\documentclass{article}
\pdfoutput=1
\usepackage{arxiv}

\usepackage[utf8]{inputenc}         % allow utf-8 input
\usepackage[T1]{fontenc}            % use 8-bit T1 fonts
\usepackage{hyperref}               % hyperlinks
\usepackage{url}                    % simple URL typesetting
\usepackage{booktabs}               % professional-quality tables
\usepackage{multirow}               % multi-row cells
\usepackage{amsfonts}               % blackboard math symbols
\usepackage{nicefrac}               % compact symbols for 1/2, etc.
\usepackage{microtype}              % microtypography
\usepackage{pdflscape}              % landscape pages (for tables)
\usepackage{listings}               % code block inclusion
\usepackage{graphicx}               % include graphics
\usepackage{subcaption}             % sub-figures
\usepackage{appendix}               % appendices
\usepackage[shortlabels]{enumitem}  % enumeration with other labels
\usepackage{threeparttable}         % Notes for tables

\usepackage[backend=biber, natbib, style=authoryear, maxcitenames=2]{biblatex}

\usepackage[dvipsnames]{xcolor}
\definecolor{codegreen}{rgb}{0,0.6,0}
\definecolor{codegray}{rgb}{0.5,0.5,0.5}
\definecolor{codepurple}{rgb}{0.58,0,0.82}
\definecolor{backcolour}{rgb}{0.95,0.95,0.92}

\usepackage{amsthm}

\usepackage{amsmath}
\usepackage{bm}
\usepackage{bbm}
\usepackage[ruled, vlined]{algorithm2e}

\newcommand{\mb}{\mathbb}
\newcommand{\mc}{\mathcal}
\renewcommand{\epsilon}{\varepsilon}

\title{Automatically Differentiable Random Coefficient Logistic Demand Estimation}

\author{
\href{https://orcid.org/0000-0001-9924-0351}{\includegraphics[scale=0.003]{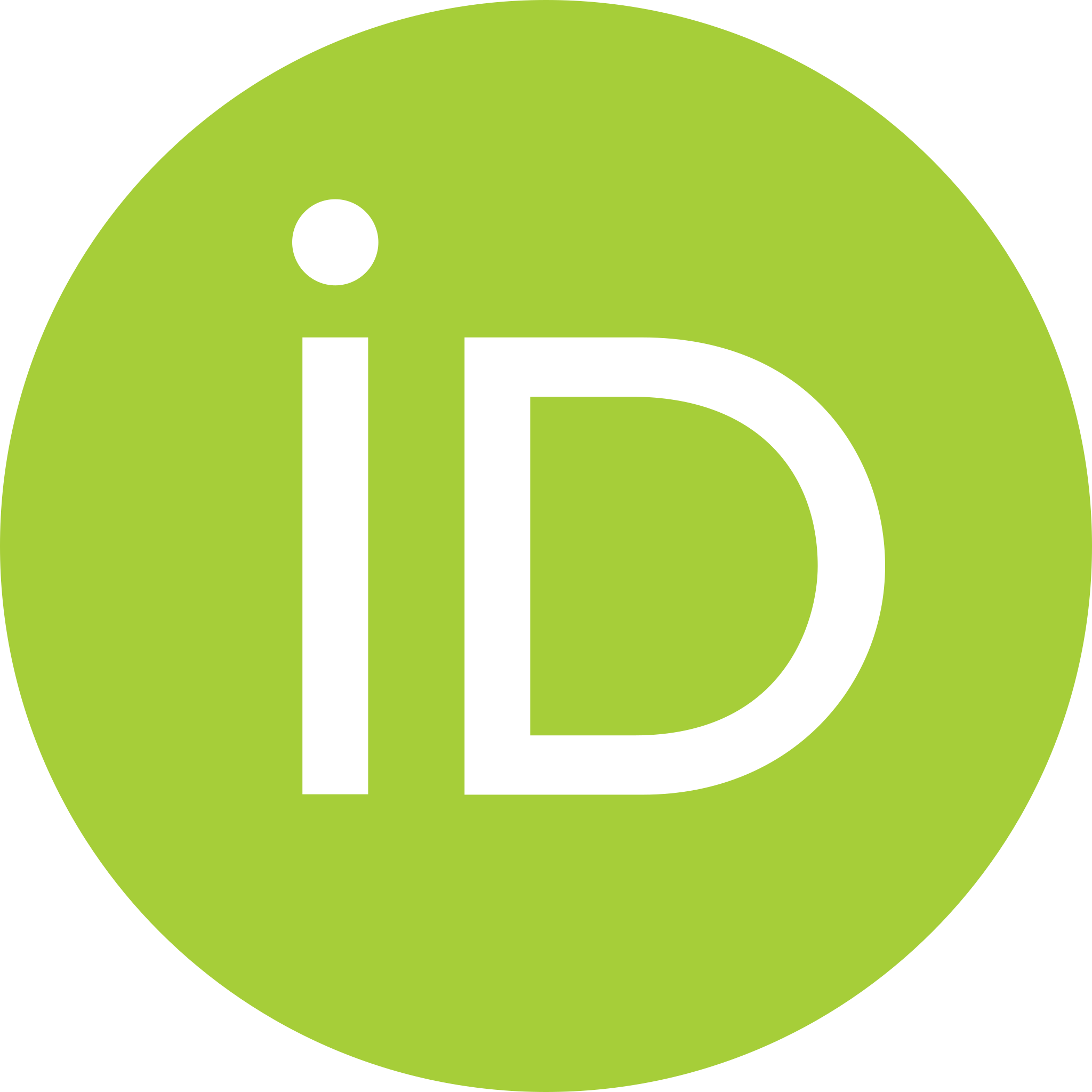}\hspace{1mm}Andrew Chia}\thanks{Thanks to Victor Chernozhukov for valuable comments.} \\
School of Engineering and Applied Sciences\\
Harvard University\\
Cambridge, MA 02138 \\
\texttt{andrewchia@g.harvard.edu} \\
}

\begin{document}

% Python Code inclusion formatting

\lstset{ %
  language=Python,                     % the language of the code
  basicstyle=\footnotesize,       % the size of the fonts that are used for the code
  numbers=left,                   % where to put the line-numbers
  numberstyle=\tiny\color{gray},  % the style that is used for the line-numbers
  stepnumber=1,                   % the step between two line-numbers. If it's 1, each line
                                  % will be numbered
  morekeywords={@fp},             % emphasize @fp decorator
  numbersep=5pt,                  % how far the line-numbers are from the code
  backgroundcolor=\color{white},  % choose the background color. You must add \usepackage{color}
  showspaces=false,               % show spaces adding particular underscores
  showstringspaces=false,         % underline spaces within strings
  showtabs=false,                 % show tabs within strings adding particular underscores
  frame=single,                   % adds a frame around the code
  rulecolor=\color{black},        % if not set, the frame-color may be changed on line-breaks within not-black text (e.g. commens (green here))
  tabsize=2,                      % sets default tabsize to 2 spaces
  captionpos=b,                   % sets the caption-position to bottom
  breaklines=true,                % sets automatic line breaking
  breakatwhitespace=false,        % sets if automatic breaks should only happen at whitespace
  title=\lstname,                 % show the filename of files included with \lstinputlisting;
                                  % also try caption instead of title
  keywordstyle=\color{codepurple},     % keyword style
  commentstyle=\color{codegreen},     % comment style
  stringstyle=\color{black},      % string literal style
}

\maketitle

\begin{abstract}
We show how the random coefficient logistic demand (BLP) model can be phrased as an automatically differentiable moment function, including the incorporation of numerical safeguards proposed in the literature. This allows gradient-based frequentist and quasi-Bayesian estimation using the Continuously Updating Estimator (CUE). Drawing from the machine learning literature, we outline hitherto under-utilized best practices in both frequentist and Bayesian estimation techniques. Our Monte Carlo experiments compare the performance of CUE, 2S-GMM, and LTE estimation. Preliminary findings indicate that the CUE estimated using LTE and frequentist optimization has a lower bias but higher MAE compared to the traditional 2-Stage GMM (2S-GMM) approach. We also find that using credible intervals from MCMC sampling for the non-linear parameters together with frequentist analytical standard errors for the concentrated out linear parameters provides empirical coverage closest to the nominal level. The accompanying \texttt{admest} \texttt{Python} package provides a platform for replication and extensibility.
\end{abstract}

% keywords can be removed

\section{Introduction}

The random coefficient logistic demand model of \citet{berry1995automobile} (henceforth BLP) has been a workhorse of the New Empirical Industrial Organization literature, allowing for varied substitution patterns across products, and accounting for endogeneity of price. The reliability of its estimation has been the subject of rigorous debate \citep{nevo2000practitioner,conlon2020best,knittel2014estimation}, and the estimator itself has been the study of many proposed advances in econometric techniques as a sophisticated yet widely used structural model \citep{hong2020blp,forneron2020inference}. 

The most common implementation of the BLP estimator involves the use of a nested fixed point (NFP) as an inner loop within an outer loop of GMM estimation, although we acknowledge the Mathematical Programming with Equilibrium Constraints (MPEC) approach of \citet{dube2012improving}, which is beyond the scope of this paper. \citet{dube2012improving,conlon2020best} find that derivative-free optimization algorithms such as the Nelder-Meade or simplex algorithms often fail to converge or converge to the wrong solution. As such, the literature has settled on the use of analytical derivatives with a derivative-based optimization algorithm such as L-BFGS. \citet{nevo2000practitioner} provides the analytical derivative for demand-only (DO) BLP in detail, and \citet{conlon2020best} indicate that the same is possible for demand-and-supply (DS) BLP, although it involves tensor products.

Notwithstanding the ongoing debate \citep{knittel2014estimation,conlon2020best} about whether the BLP estimator can be reliably implemented, it is clear that gradient-based estimation with tight tolerances in the NFP are required for any hope of reliability for the non-linear, non-convex optimization problem. However, the reliance on analytical derivatives poses challenges for the robust and reliable estimation of variants on the BLP model such as dynamic demand decisions involving NFPs within NFPs \citep{hendel2006measuring} or sequential NFPs \citet{lee2013vertical}. Quasi-Bayesian estimation using the Laplace Transform Estimator (LTE) of \citet{chernozhukov2003mcmc} with the latest gradient-based Markov Chain Monte Carlo (MCMC) samplers such as Hamiltonian Monte Carlo (HMC) is also difficult in practice. By contrast, automatic differentiation offers easy extensibility, since automatically differentiable modifications preserve the automatic differentiability of the resulting objective function, and straightforward integration into the existing software ecosystem around the latest MCMC sampling techniques.

Our contribution is fourfold. First, we derive the BLP objective function (both DO and DS variants) using block diagonal matrix multiplication in place of for-loops so that the resulting objective function is reverse mode automatically differentiable. Second, we briefly outline what automatic differentiation is, and how the BLP objective function can be phrased as an easily extensible composition of automatically differentiable building blocks. Third, we point out the algorithmic improvements in our reference implementation, the \texttt{admest}\footnote{\textbf{A}utomatically \textbf{D}ifferentiable \textbf{M}-\textbf{EST}imation.} \texttt{Python} package, including where we depart from standard practice in the literature. Fourth, we evaluate these algorithmic improvements in Monte Carlo simulations examining the relative performance of 2S-GMM, CUE, and LTE estimators with automatic differentiation, with 2S-GMM using analytical gradients as a benchmark.

Preliminary simulation results confirm previous theory \citep{hansen1996finite,newey2004higher,newey2009generalized,donald2000jackknife} that the CUE has a lower bias but higher variance compared to the 2S-GMM estimator given the weak instrument setting of the Monte Carlo simulation. On the performance of the LTE, we find that lifting the credible intervals from the MCMC sampler directly achieves good empirical coverage for the parameters estimated in the outer GMM loop. This is in contrast to \citet{hong2020blp}, who found it necessary to correct for simulation error to achieve good empirical coverage. Using credible intervals directly results in poor coverage of the concentrated out linear parameters, but using the LTE posterior mean as a point estimate for frequentist uncertainty quantification on these parameters restores good empirical coverage compared to frequentist estimators. Our gradient-based No-U-Turn Sampler variant of HMC (NUTS-HMC) delivers higher effective sample sizes than a Random Walk Metropolis-Hastings (RWMH) implementation with significantly fewer draws \citep{forneron2020inference}.

In Section \ref{sec:model} we review the BLP model and show how to construct the the Continuously Updating Estimator (CUE) of \citet{hansen1996finite}. Section \ref{sec:autodiff} gives a brief overview of automatic differentiation, and how it can be applied to the BLP model. Section \ref{sec:algo_improv} details the algorithmic improvements of our choice of frequentist and Bayesian estimation techniques. Section \ref{sec:mc_setup} documents the set up of our Monte Carlo simulation exercise, while Section \ref{sec:mc_results} contains the result of our empirical exercise. Section \ref{sec:conclude} concludes. Appendix \ref{app:do2s} derives the 2S-GMM objective function and the analytical gradient without the use of for-loops.

\section{BLP Model and Computation}\label{sec:model}

\subsection{Demand-Only Model and Setup}\label{sec:blpdo}
As in \citet{berry1995automobile}, we start with some individual $i$ in market $t$ that faces a choices of $J_t$ products, including the outside alternative denoted $j=0$. The utility of consumption for each product $u_{ijt}$ is given by

$$ u_{ijt} = \sum_{k\in K_1}x_k\beta_{k,r} - \alpha_r p_{jt} + \xi_{jt} +  \varepsilon_{ijt} $$

where the parameter vector $(\beta_r, -\alpha_r)'$ has a mean component $(\beta, -\alpha)'$ and covariance matrix with Cholesky decomposition $\Sigma$, $\varepsilon_{ijt}$ is a Type 1 Extreme Value error, and $\xi_{jt}$ is an unobserved structural error. Under the \citet{berry1995automobile} formulation, $\Sigma$ is assumed to be a diagonal as opposed to a lower triangular matrix, with the non-zero components denoted $\sigma_k$. Define $\theta_1 = (\beta, -\alpha)$, the full vector of features as $X_{1,jt}'$, and the subset of features with non-zero entries in the covariance matrix as $X_{2,jt}'$, and the standardized vector of individual-level taste heterogeneity as $\nu_i$, then we have

\begin{align*}
u_{ijt} &= \delta_{jt} + \mu_{ijt} + \varepsilon_{ijt} \\
\delta_{jt} &= X_{2,jt}'\theta_1 + \xi_{jt}\\
\mu_{ijt} &= X_{2,jt}'\Sigma\nu_i = \sum_{k=1}^{K_2}x_{2,jt,k}\sigma_k\nu_{i,k}
\end{align*}

Consumers select the product that gives them the most utility, where the value of the outside good is normalized to zero so $u_{i0t} = 0\ \forall\ i,t$. Let $\mc{N}(t)$ denote the set of products in market $t$, then the observable market shares are given by

$$ s_{jt} = \int \frac{\exp(\delta_{jt} + \mu_{ijt})}{1 + \sum_{k \in \mc{N}(t)}\exp(\delta_{jt} + \mu_{kjt})} dF(\nu)$$

In practice, this integral is computed numerically using simulation under a distributional assumption for $\nu_i$. We assume independent normality (across products and individuals), as is common in the literature. \citet{berry1995automobile} show that we can invert the simulated and observed shares to recover $\delta_{jt}$ using a contraction mapping. Once $\delta_{jt}$ is recovered, we can retrieve the structural error $\xi_{jt}$. Denote all observable data by $D$. With demand-side instruments $Z^D$, we have the moment conditions $g(\theta; D_{jt}) = Z^{D}_{jt}\xi_{jt}$, and $\mb{E}\left[g(\theta; D_{jt})\right] = 0$, which we can then estimate using GMM.

\subsection{Constructing the Demand-Only BLP Objective Function}\label{sec:blpdo_obj}
The following derivation is novel in that it eschews the use of for-loops while being numerically equivalent, avoiding the criticism \citep{conlon2020best} of previous vectorization methods \citep{nevo2000practitioner,knittel2014estimation}. Since each iteration of the for-loop across markets does not depend on computations from other iterations, the loop constitutes a finite-dimensional linear operator which definitionally has an equivalent expression as a series of matrix operations -- in this case, appropriately chosen block diagonal matrices.

In considering numerical integration, we make the subjective choice of assuming overlapping simulation draws \citep{hong2020blp}, although there are potential efficiency gains from using separate draws for each market \citep{kristensen2017higher,lee1995asymptotic}. We further use quasi-Monte Carlo nodes with equal weight rather than quadrature nodes, although the derivations easily extend to the case of quadrature nodes.

\paragraph{Share Inversion}\mbox{}

The contraction map given by \citet{berry1995automobile} is

$$ \delta_{n+1}(\delta_n; \hat{s}(\delta_n; \theta_2), \mc{S}) = \delta_n + \log(\mc{S}) - \log(\hat{s}(\delta_n; \theta_2)) $$

where $\mc{S}$ is the observed share (and its dependence is henceforth suppressed for notational simplicity), and $\hat{s}$ is the simulated shares. For a given product $j$ in market $t$, with each simulated individual denoted by $i$, the simulated share is given by

$$ \hat{s}_{ji}(\delta_n; \theta_2) = \frac{\exp(\delta_n + \mu_{ji}(\theta_2))}{1 + \sum_j \exp(\delta_n + \mu_{ji}(\theta_2))} $$
$$ \mu_{ji}(\theta_2) = \sum_kx_{j,k}\nu_{i,k}\theta_{2,k} $$

At each iteration of the fixed point the term $\mu_i(\theta_2)$ does not change (this value is fixed in the outer loop of the non-linear GMM algorithm). Therefore, it makes sense to cache the values of $\mu_{ji}$ for re-use across each fixed point iteration, rather than computing it at every iteration. Note also that we can express the $\mu$ matrix for all goods and individuals as

$$ \mu(\theta_2) = X_2 \Omega_{\theta_2} \nu' \in \mb{R}^{J \times R}$$

where $\nu \in \mb{R}^{R\times K_2}$ is the matrix of simulation draws and $\Omega_{\theta_2}$ is a diagonal matrix with diagonal entries corresponding to $\theta_2$. The compute time of this $\mu$ matrix scales with the number of simulation draws $R$, so caching it yields significant savings in compute time especially when we increase $R$ for more precision in numerical integration. 

To think about how to recover the shares across multiple markets in a single matrix operation, we first examine the case of a single market with number of products $J_t$, where we want to recover the matrix $\hat{s}(\delta_n;\theta_2) \in \mb{R}^{J_t \times R}$. We already have the matrix $\mu(\theta_2) \in \mb{R}^{J_t\times R}$, and we can perform column-wise addition (denoted $+^c$) of $\delta_n$ on it and apply the exponential function element-wise to get $\exp(\delta_n + \mu_{ji}(\theta_2))$, so the only term that remains is $\sum_j \exp(\delta_n + \mu_{ji}(\theta_2))$, which we want to compute at the individual level (so each entry within the same column has the same value, and this value should differ across the $R$ columns). Taking $\iota_{J_t} \in \mb{R}^{J_t\times J_t}$ (the ``summation matrix'') to be the matrix of all ones, we can use the following formula

$$ \iota_{J_t} \exp(\delta_n +^c \mu(\theta_2)) \in \mb{R}^{J_t\times R} $$

This gives a single expression for the matrix of simulated shares as

$$ \hat{s}(\delta_n;\mu(\theta_2)) = \exp(\delta_n +^c \mu(\theta_2)) \oslash \left[1 + \iota_{J_t}\exp(\delta_n +^c \mu(\theta_2))\right] $$

where the addition of 1 is element-wise and $\oslash$ denotes element-wise division (as opposed to the Hadamard element-wise product denoted $\odot$). Moving to the many markets scenario, define instead $\iota_J$ to be the block diagonal with $\iota_{J_t}$ for each market $t$ in each block\footnote{The matrix $\iota_J$ is also constant across iterations of the fixed point, so it can be pre-computed and cached. In fact, it is invariant to the choice of $\theta_2$, so it should be passed in from outside of the outer non-linear GMM loop to save even more time.}, so we have

$$ \iota_J = 
\begin{bmatrix}
    \iota_{J_1} & 0           & \cdots & 0 \\ 
    0           & \iota_{J_2} & \cdots & 0 \\
    0           & 0           & \ddots & 0 \\
    0           & 0           & \cdots & \iota_{J_T} 
\end{bmatrix} $$

Then the share matrix across all products, markets, and individuals can be computed by

$$ \mb{R}^{J \times R} \ni \hat{s}(\delta_n;\theta_2) = \exp(\delta_n +^c \mu(\theta_2)) \oslash \left[1 + \iota_J\exp(\delta_n +^c \mu(\theta_2))\right] $$

\paragraph{Numerical Safeguards}\mbox{}

\citet{conlon2020best} note that in situations where $\delta_n + \mu_{ji}(\theta_2) \gg 0$, $\exp(\delta_n + \mu_{ji(\theta_2)})$ will overflow, leading to numerical issues when trying to divide infinity by infinity. They propose choosing (at the market level) $m_i = \max\{0, \max_{j}\delta_n + \mu_{ji}(\theta_2)\}$, and then note that

$$ \frac{\exp(\delta_n + \mu_{ji}(\theta_2))}{1 + \sum_j \exp(\delta_n + \mu_{ji}(\theta_2))} = \frac{\exp(\delta_n + \mu_{ji}(\theta_2) - m_i)}{\exp(-m_i) + \sum_j \exp(\delta_n + \mu_{ji}(\theta_2) - m_i)} $$

The choice of maximum provides the best protection against numerical overflow, but it is not expressible as a matrix operation. We propose re-centering using the mean instead. This guards against overflow and underflow, is expressible as a matrix operation, and is likely to be sufficient as long as $\|\theta_2\|_\infty$ is of reasonable size. The failure mode when $\|\theta_2\|_\infty \gg 0$ is unlikely to be encountered in practice, as this would lead to other numerical issues in the objective function.\footnote{\citet{conlon2020best} recommend using bounds on the value of $\theta_2$ to ensure a well-behaved objective function. Within reasonable bounds, the performance of the mean is likely to be at least equivalent to using the max.} This operation is in fact similar to the summation within markets at the individual level that we have already detailed. Define the ``averaging matrix''\footnote{One can retrieve $\iota_J$ from $\iota_{J/J}$ by performing a boolean check for non-zero elements (encoding true as 1 and false as 0), so in implementation one should construct $\iota_{J/J}$ first, and then perform the boolean check to retrieve $\iota_J$, with both operations happening outside the BLP algorithm.}

$$ \iota_{J/J} = 
\begin{bmatrix}
    \iota_{J_1}/J_1 & 0           & \cdots & 0 \\ 
    0           & \iota_{J_2}/J_2 & \cdots & 0 \\
    0           & 0           & \ddots & 0 \\
    0           & 0           & \cdots & \iota_{J_T}/J_T
\end{bmatrix} $$

then the matrix given by $\iota_{J/J} (\delta_n +^c \mu(\theta_2))$ gives the market level mean of $\delta_n +^c \mu(\theta_2)$ for a given individual. Incorporating this guard against numerical overflow, our final expression for the shares is given by

$$ \hat{s}(\delta_n;\theta_2) = \exp(\delta_n +^c \mu(\theta_2) - \iota_{J/J} [\delta_n +^c \mu(\theta_2)]) \oslash \left[1 + \iota_J\exp(\delta_n +^c \mu(\theta_2) - \iota_{J/J} [\delta_n +^c \mu(\theta_2)])\right] $$

To take the average over individuals, we can just take the row sums of this matrix.\footnote{It is advisable to cache $\hat{s}(\delta_n;\theta_2)$ before row-summation, as it will be needed in this form for other computations.} We are now ready to construct the CUE objective function for the BLP problem.

\paragraph{Demand-Only CUE}\mbox{}

We partition the parameter space $\theta$ into two components. $\theta_1 \in \mb{R}^{K_1}$ is the vector of linear parameters with the corresponding design matrix $X_1$, while $\theta_2 \in \mb{R}^{K_2}$ is the vector of non-linear parameters governing heterogeneous tastes with the corresponding design matrix $X_2$. We denote the fixed point solution of the contraction mapping $\lim_{n\to\infty}\delta_n(\delta_0; \mu(\theta_2)) := \lim_{n\to\infty}\delta_n(\delta_{n-1}; \mu(\theta_2))$, and the matrix of instruments by $Z\equiv Z^D$. The Nested Fixed Point (NFP) algorithm of \citet{berry1995automobile} implemented as a CUE objective function is given by Algorithm \ref{alg:docue}.\footnote{For comparison and completeness of documentation, the 2S-GMM version of demand-only NFP BLP along with its gradient without for-loops is given in Appendix \ref{app:do2s}.}\footnote{We scale the objective function by $\frac{1}{2}$ to be consistent with the LTE formulation of \citet{chernozhukov2003mcmc}.}

There are two remarks to be made about the formulation. First, while the rest of the literature uses a one-step linear GMM to compute the concentrated out parameters $\theta_1$, we use a 2-Stage GMM (2S-GMM) instead, since we do not have the weight matrix as a parameter.\footnote{The weight matrix is initialized using the 2-Stage Least Squares (2SLS) weight matrix $(Z'Z)^{-1}$.} Second, for the computation of the weight matrix within the CUE objective function, standard errors, and computing out concentrated out parameters we need an estimator for $Var(Z'\xi)$. Under the assumption of heteroskedascity, one popular choice is $\hat{Var}(Z'\xi) = Z'\Omega_\xi Z$ where $\Omega_\xi$ is a diagonal matrix with the squared residuals $\xi_{jt}^2$ as its diagonal entries. We find better performance using the following formulation as proposed by \citet{gallant2007statistical}

$$\hat{Var}(Z'\xi) = Z'\Omega_\xi Z - Z'\xi\xi'Z $$

\begin{algorithm}
\DontPrintSemicolon
\KwData{$X_1 \in \mb{R}^{J\times K_1},\ X_2 \in \mb{R}^{J\times K_2}$ the design matrices of features that enter linearly and non-linearly respectively; $Z \in \mb{R}^{J \times z}$ the matrix of instruments; $\nu \in \mb{R}^{R \times K_2}$ the matrix of simulation draws; $\mc{S} \in\mb{R}^J$ the vector of observed shares; $\iota_{J/J}, \iota_J$ as in Section \ref{sec:blpdo_obj}.}
\KwIn{The value of $\theta_2$ being considered, the starting point for the fixed point equation $\delta_0$.}
\KwResult{The CUE objective function $q(\theta_2)$}
\Begin{
    $\mb{R}^{J\times R} \ni \mu(\theta_2) \longleftarrow X_2 \Omega_{\theta_2} \nu'$\;
    $\mb{R}^J \ni \hat{\delta} \longleftarrow \lim_{n\to\infty}\delta_n(\delta_0; \mu(\theta_2))$\;\;

    \textit{Use 2S-GMM to compute $\theta_1$}\;
    $\mb{R}^{z \times z} \ni W^{(1)} = (Z'Z)^{-1}$\;
    $\mb{R}^{K_1} \ni \hat{\theta_1}^{(1)} \longleftarrow \left[X_1'ZW^{(1)} Z'X_1\right]^{-1}X_1'ZW^{(1)} Z'\hat{\delta}$\;
    \textit{Estimate the Second Stage Weight Matrix}\;
    $\mb{R}^J \ni \xi^{(1)} \longleftarrow \hat{\delta} - X_1\hat{\theta}_1^{(1)} $\;
    $\mb{R}^{z \times z} \ni W^{(2)} = (Z'\Omega_{\xi^{(1)}}Z - Z'\xi^{(1)}\xi^{(1)'}Z)^{-1}$\;
    $\mb{R}^{K_1} \ni \hat{\theta_1}^{(2)} \longleftarrow \left[X_1'ZW^{(2)}Z'X_1\right]^{-1}X_1'ZW^{(2)}Z'\hat{\delta}$\;
    $\mb{R}^J \ni \xi^{(2)} \longleftarrow \hat{\delta} - X_1\hat{\theta}_1^{(2)}$\;\;

    \textit{Estimate the weight matrix for the overall objective}\;
    $\mb{R}^{z \times z} \ni W = (Z'\Omega_{\xi^{(2)}}Z - Z'\xi^{(2)}\xi^{(2)'}Z)^{-1}$\;
    $q(\theta_2) \longleftarrow \frac{1}{2}\xi^{(2)'} Z W Z'\xi^{(2)} $
}
\caption{NFP GMM Objective (Demand-Only)}
\label{alg:docue}
\end{algorithm}

We use the standard GMM standard errors\footnote{We acknowledge the work of \citet{newey2009generalized} in showing that the usual GMM standard errors are biased downwards for the CUE.} for $\hat{\theta}$ denoted $\Sigma_{\hat{\theta}}$, which \citet{berry1995automobile} derive to be

$$ \Sigma_{\hat{\theta}} = \left[\frac{\partial \xi(\theta_2)'Z}{\partial \theta} \left[\sum_{i=1}^3 V_i\right]^{-1} \frac{\partial Z'\xi(\theta_2)}{\partial \theta}\right]^{-1} $$

where $V_1,\ V_2,\ V_3$ represent the variation arising from the process generating the product characteristics, from the consumer sampling process, and from the simulation process respectively. We have $V_1 = Var(Z'\xi)$, and with a large number of underlying consumers, $V_2$ is negligible. With a sufficient number of simulation draws relative to the number of products, $V_3$ is also negligible, so we use as an empirical estimator 

$$ \Sigma_{\hat{\theta}} = \left[\left(\frac{\partial \xi(\theta_2)'Z}{\partial \theta}\right) \left(Z'\Omega_\xi Z - Z'\xi\xi'Z\right)^{-1} \frac{\partial Z'\xi(\theta_2)}{\partial \theta}\right]^{-1} $$

evaluated at the point estimate $\hat{\theta}$, where $\frac{\partial Z'\xi(\theta_2)}{\partial \theta}$ is computed using automatic differentiation.

\subsection{Incorporating Supply-Side into the Model}
\citet{conlon2020best} show that as long as the supply side is correctly specified, its inclusion can increase the efficiency of estimation, especially for situations where the excluded demand-side instruments are relatively weak. Using the first order condition of firms under the assumption of Bertrand competition, we can generate an additional set of moment conditions. At the market $t$ level, we have the multi-product Bertrand markup $\eta_t(p_t,s_t,\theta_2)$ estimated by

$$ \eta_t(p_t, s_t, \theta_2) \equiv \Delta_t(p_t)^{-1}s_t = p_t - c_t $$

where $\Delta_t(p_t,s_t,\theta_2)$ is the intra-firm demand derivative given by

$$ \Delta_t(p_t) \equiv -\mc{H}_t \odot \frac{\partial s_t}{\partial p_t}(p_t) $$

$\mc{H}_t$ refers to the ownership matrix with each $(i,j)$ entry indicating whether the same firm produces products $i$ and $j$. With this, we can estimate marginal costs by taking $c_t = p_t - \eta_t(\theta_2)$, and parametrize it (across all markets and products) as

$$ f_{MC}(p - \eta(\theta_2)) = X_3\gamma + \omega$$

where $\omega$ is a structural error term, $\gamma$ is the vector of coefficients on the linear supply-side features, and $X_3$ is a design matrix of features that enter supply linearly. $f_{MC}$ is the functional form assumed for marginal costs, typically the identity function or the logarithm. This allows the construction of moment conditions $\mb{E}\left[\omega_{jt}Z^S_{jt}\right] = 0$. We can then estimate the full demand and supply system with multiple equation GMM.

\subsection{Estimating Demand-Supply BLP}
\paragraph{Empirically Estimating Marginal Cost}\mbox{}

We focus on an empirical estimate for $\frac{\partial s_t}{\partial p_t}$. Denote the coefficient on price as $\alpha$.\footnote{Our notation here differs from the rest of the literature in that $\alpha < 0$ is the coefficient on price rather than its size.}, and $\alpha_i = \alpha + \theta_{2,\text{price}} \nu_{i,\text{price}}\ \forall\ i = 1,\ldots, R$ to be the vector of (simulated) individual level coefficients on price, and the matrix $\Omega_\alpha$ to be a diagonal matrix with $\alpha_i$ as its diagonal elements. In the case where price is not a non-linear feature, set $\theta_{2,\text{price}} = 0$ and this reduces to $\alpha_i = \alpha\ \forall\ i$. Then at the market $t$ level we have

$$ \frac{\partial s_j}{\partial p_k} = \begin{cases}
-\int \alpha_i \hat{s}_{ij}(1-\hat{s}_{ik})d(\nu)\ \forall\ j=k \\
\int \alpha_i \hat{s}_{ij}\hat{s}_{ik}d(\nu)\ \forall\ j\neq k
\end{cases}$$

Following \citet{morrow2011fixed}, we can decompose this into a diagonal matrix $\Lambda_t$ and a dense matrix $\Gamma_t$ so we have

\begin{align*}
\Delta_t &= \Lambda_t - \Gamma_t \\
diag(\Lambda_t) &= rowMeans(\hat{s}_t\Omega_\alpha) \\
\Gamma_t &= \hat{s}_t\Omega_\alpha\hat{s}_t/R
\end{align*}

To move to the many markets scenario without for-loops, we note that we can simply substitute the problem-level matrices for $\Lambda$. For $\Gamma$, we want a block diagonal matrix rather than a dense matrix. Hence, we ``over-provision'' a dense matrix and use Hadamard multiplication by $\iota_J$ to ``slice'' out the entries that we want\footnote{See Appendix \ref{app:do2s} for a similar approach taken for the analytical gradient, and associated trade-offs.}, so we have

$$ \Gamma = \hat{s}\Omega_{\alpha}\hat{s}' \odot \iota_j / R$$

We then estimate $\hat{\eta}$ with\footnote{We use the Moore-Penrose Pseudo-Inverse denoted $\dagger$ when inverting the matrix of intra-firm derivatives because the matrix can be computationally singular when the observed outside share is large and predicted shares are small.}

$$ \hat{\eta} = \left(-\mc{H}\odot \frac{\partial \hat{s}}{\partial p}\right)^\dagger \mc{S} $$

\paragraph{Demand-Supply CUE}\mbox{}

We partition the parameter space $\theta = (\theta_1, \theta_2, \theta_3)$, where $\theta_1 \in \mb{R}^{K_1}$ are the parameters on the features that enter demand linearly including the price parameter $\alpha$ (which is negative), $\theta_2 \in \mb{R}^{K_2}$ are the parameters on the features that enter demand non-linearly, and $\theta_3 \in \mb{R}^{K_3}$ are the parameters on the features that enter supply linearly. We present the DS CUE objective function where the outer loop only comprises the non-linear parameters $\theta_2$. This is a departure from \citet{conlon2020best}, who include the linear parameter on price $\alpha$ in the outer GMM loop. They do this because it allows for the sharing of the weight matrix between the concentrated out parameters and the outer GMM loop. Our simulation results suggest that unless one imposes parameter bounds, this tends to have poor performance when applied to the CUE. Instead, we estimate the demand-side of BLP, using a 2-Stage linear IV GMM procedure to recover the concentrated out linear parameters including $\hat{\alpha}$, and then estimate the supply-side using $\hat{\alpha}$. Doing so forces $\hat{\alpha}$ to be in a feasible region (and in particular, negative) without the use of any parameter bounds.

On top of the data used for demand-only BLP, we need $\mc{H}$ the conduct matrix, $f_{MC}$ the functional form for marginal costs, $X_3$ the matrix of features that enter supply linearly, and $Z^S$ supply-side instruments, the last of which is used to construct $Z \equiv \begin{bmatrix} Z^D & 0 \\ 0 & Z^S \end{bmatrix}$ the full block matrix of instruments. Denote the vector of firm indices to be $\vec{firm}$, then one trick to computing $\mc{H}$ is to compute the boolean matrix

$$ \mc{H}: \left[\vec{firm}(1/\vec{firm})'\right] \overset{?}{=} 1 $$

encoding \texttt{True} to be 1 and \texttt{False} to be 0. The CUE objective function is given by Algorithm \ref{alg:sscue}, and uncertainty quantification proceeds in similar fashion to demand-only BLP. Denote $\varepsilon = (\xi, \omega)$, then we have

$$ \Sigma_{\theta} = \left[\frac{\partial \varepsilon(\theta)'Z}{\partial \theta} \left(Z'\Omega_\varepsilon  Z - Z'\varepsilon\varepsilon'Z\right)^{-1} \frac{\partial Z'\varepsilon(\theta)}{\partial \theta}\right]^{-1} $$

where $\frac{\partial Z'\varepsilon(\theta_2)}{\partial \theta}$ is computed using automatic differentiation.

\begin{algorithm}
\DontPrintSemicolon
\KwData{$X_1 \in \mb{R}^{J\times K_1},\ X_2 \in \mb{R}^{J\times K_2}, X_3 \in \mb{R}^{J\times K_3}$ the design matrices of features that enter demand linearly, demand non-linearly, and supply linearly respectively; $p \in \mb{R}^J$ the vector of prices; $\mc{H} \in \mb{R}^{J\times J}$ the conduct matrix, $Z^D \in \mb{R}^{J \times z_d},\ Z^S \in \mb{R}^{J\times z_s}$ the demand and supply instrument matrices respectively; $\nu \in \mb{R}^{R \times K_2}, \mc{S} \in\mb{R}^J, \iota_{J/J}, \iota_J$ as in Algorithm \ref{alg:docue}.}
\KwIn{The value of $\theta_2$ being considered, the starting point for the fixed point equation $\delta_0$.}
\KwResult{The CUE objective function $q(\theta_2, \alpha)$}
\Begin{
    \textit{Demand-Side Estimation}\;
    $\mb{R}^{J\times R} \ni \mu(\theta_2) \longleftarrow X_2 \Omega_{\theta_2} \nu'$\;
    $\mb{R}^J \ni \hat{\delta} \longleftarrow \lim_{n\to\infty}\delta_n(\delta_0; \mu(\theta_2))$\;
    $\mb{R}^{J \times R} \ni \hat{s}(\hat{\delta};\theta_2) \longleftarrow \exp(\hat{\delta} +^c \mu(\theta_2)) \oslash \left[1 + \iota_J\exp(\hat{\delta} +^c \mu(\theta_2))\right]$\;\;

    \textit{Use 2S-GMM to compute $\theta_1$}\;
    $\mb{R}^{z_d \times z_d} \ni W^{(1)} = (Z^{D'}Z^D)^{-1}$\;
    $\mb{R}^{K_1} \ni \hat{\theta_1}^{(1)} \longleftarrow \left[X_1'Z^DW^{(1)} Z^{D'}X_1\right]^{-1}X_1'Z^DW^{(1)} Z^{D'}\hat{\delta}$\;
    \textit{Estimate the Second Stage Weight Matrix}\;
    $\mb{R}^J \ni \xi^{(1)} \longleftarrow \hat{\delta} - X_1\hat{\theta}_1^{(1)} $\;
    $\mb{R}^{z_d \times z_d} \ni W^{(2)} = (Z^{D'}\Omega_{\xi^{(1)}}Z^D - Z^{D'}\xi^{(1)}\xi^{(1)'}Z^D)^{-1}$\;
    $\mb{R}^{K_1} \ni \hat{\theta_1}^{(2)} \longleftarrow \left[X_1'Z^DW^{(2)}Z^{D'}X_1\right]^{-1}X_1'Z^DW^{(2)}Z^{D'}\hat{\delta}$\;
    $\mb{R}^J \ni \xi^{(2)} \longleftarrow \hat{\delta} - X_1\hat{\theta}_1^{(2)}$\;\;

    \textit{Supply-Side Estimation}\;
    \textit{Estimate marginal costs using $\hat{\alpha}$ from previous step}\;
    $\mb{R}^{J\times J} \ni \Lambda_{ii} \longleftarrow rowMeans(\hat{s} \Omega_{\hat{\alpha}})$\;
    $\mb{R}^{J\times J} \ni \Gamma \longleftarrow \hat{s}\Omega_{\hat{\alpha}}\hat{s}' \odot \iota_j / R$\;
    $\mb{R}^{J\times J} \ni \frac{\partial\hat{s}}{\partial p} \longleftarrow \Lambda - \Gamma$\;
    $\mb{R}^J \ni \hat{\eta} \longleftarrow \left(-\mc{H} \odot \frac{\partial\hat{s}}{\partial p}\right)^{\dagger}\mc{S}$\;
    $\mb{R}^J \ni \hat{c} \longleftarrow f_{MC}(\hat{\eta} - p)$\;\;

    \textit{Use 2S-GMM to compute $\theta_3$}\;
    $\mb{R}^{z_s \times z_s} \ni W^{(1)} = (Z^{S'}Z^S)^{-1}$\;
    $\mb{R}^{K_3} \ni \hat{\theta_3}^{(1)} \longleftarrow \left[X_3'Z^SW^{(1)} Z^{S'}X_3\right]^{-1}X_3'Z^SW^{(1)} Z^{S'}\hat{c}$\;
    \textit{Estimate the Second Stage Weight Matrix}\;
    $\mb{R}^J \ni \omega^{(1)} \longleftarrow \hat{c} - X_3\hat{\theta}_3^{(1)} $\;
    $\mb{R}^{z_s \times z_s} \ni W^{(2)} = (Z^{S'}\Omega_{\omega^{(1)}}Z^S - Z^{S'}\omega^{(1)}\omega^{(1)'}Z^S)^{-1}$\;
    $\mb{R}^{K_3} \ni \hat{\theta_3}^{(2)} \longleftarrow \left[X_3'Z^SW^{(2)}Z^{S'}X_3\right]^{-1}X_3'Z^SW^{(2)}Z^{S'}\hat{c}$\;
    $\mb{R}^J \ni \omega^{(2)} \longleftarrow \hat{c} - X_3\hat{\theta}_3^{(2)}$\;\;

    \textit{Pool residuals and instruments from demand and supply}\;
    $\mb{R}^{2J} \ni \hat{\epsilon} \longleftarrow (\hat{\xi}, \hat{\omega})$\;
    $\mb{R}^{2J \times (z_d + z_s)} \ni Z \longleftarrow \begin{bmatrix} Z^D & 0 \\ 0 & Z^S \end{bmatrix}$\;
    \textit{Estimate the weight matrix for the overall objective}\;
    $\mb{R}^{(z_d + z_s) \times (z_d + z_s)} \ni W = (Z'\Omega_{\hat{\epsilon}^{(2)}}Z - Z'\varepsilon\varepsilon'Z)^{-1}$\;
    $q(\theta_2, \alpha) \longleftarrow \frac{1}{2}\hat{\epsilon}^{(2)'} Z W Z'\hat{\epsilon}^{(2)} $
}
\caption{NFP GMM Objective (Demand and Supply)}
\label{alg:sscue}
\end{algorithm}

\section{Applying Automatic Differentiation}\label{sec:autodiff}

Traditionally, gradient-based estimation of the BLP objective function has relied on deriving the analytical gradient. However, \citet{conlon2020best} point out that the mere inclusion of a supply-side greatly complicates this endeavour, requiring the use of tensor products. Automatic differentiation offers a way to estimate gradients up to numerical precision (avoiding the error from a finite differences approximation) without the need for extensive derivation. 

\subsection{What is Automatic Differentiation?}
This section is not a substitute for a more extensive treatment of automatic differentiation found in \citet{rall1996introduction,jax2018github}. Readers interested in the history of automatic differentiation and its use in machine learning can refer to \citet{iri1991history} and \citet{baydin2018automatic} respectively. Given a function $f: \mb{R}^n \to \mb{R}^m$, we think of the Jacobian evaluated at $x \in \mb{R}^n$ denoted $\partial f(x)$ to be an $m\times n$ matrix representing the differential of $f$ at every point where $f$ is differentiable. For some $y \in \mb{R}^m$, the Jacobian-vector product $\partial f(x)\cdot y$ represents the best linear approximation to the change $y-x$ in the neighborhood around $x$. \textbf{Forward Mode} automatic differentiation computes the Jacobian-vector product. The Jacobian of the function is build one column at a time by keeping track of the partial derivative of each intermediate step in the function, or the tangent of the primal. This is the most efficient for the case where $f: \mb{R} \to \mb{R}^m$, where the entire Jacobian matrix is computed in a single pass. \textbf{Reverse Mode} automatic differentiation computes the vector-Jacobian product, which is the transpose of the Jacobian vector product. The Jacobian matrix is computed one row at a time in two sweeps. The first forward sweep keeps track of the dependencies that intermediate values have on the original input, and then the reverse sweep computes the derivative by propagating the adjoints in reverse. The special case where $f: \mb{R}^n \to \mb{R}$, where the Jacobian is computed in a single pass, is known as \textbf{back-propagation}. Given a scalar-valued objective function, back-propagation is the most efficient of the two automatic differentiation modes, explaining its popularity in the machine learning literature and why it is our mode of choice.

Similar to the chain rule for analytical derivatives, compositions of automatically differentiable functions preserve automatic differentiability. This means that we get gradients of any variant of the original BLP algorithm ``for free'', provided the variation is phrased as an automatically differentiable function. The simplest example of this is that any automatically differentiable moment function admits an automatically differentiable CUE objective function, since the weight matrix estimation step only involves matrix operations.

\subsection{What is Automatically Differentiable?}
With modern machine learning libraries, compositions of the following building blocks are automatically differentiable:
\begin{enumerate}
    \item Matrix operations, including non-linear operations such as matrix inversion.
    \item Elementary functions such as trigonometric functions, taking the exponential or logarithm.
    \item Indicator functions.
    \item Fixed-Point solutions.
\end{enumerate}

By contrast, automatic differentiation does not work well with loop constructs, both for-loops and while-loops. Forward-mode differentiation is possible with loop-constructs by ``unrolling'' the loops at every evaluation of the function, a costly operation, but reverse-mode differentiation, and therefore back-propagation, is not possible. It is clear that the derivation of both the DO and DS BLP objective functions in Section \ref{sec:model} only comprises compositions of building blocks 1.--4., and is thus automatically differentiable. The derivative of a fixed-point solution relies on the implicit function theorem, and is typically not supported out of the box by most major automatic differentiation libraries. Hence, we detail our implementation of the reverse-mode automatic derivative of a fixed-point solution.

\paragraph{Reverse Mode Automatic Derivative of Fixed-Point Solutions}\mbox{}

Numerical methods such as fixed-point iteration that effectively rely on a while-loop are not amenable to back-propagation, and are slow to automatically differentiate using forward mode. However, \citet{nevo2000practitioner} already observed that we can use the implicit function theorem to retrieve the derivative of $\delta$ with respect to $\theta_2$ given by

$$ \frac{\partial\delta}{\partial\theta_2} = -\left[\frac{\partial \hat{s}}{\partial\delta}\right]^{-1} \frac{\partial \hat{s}}{\partial\theta_2} $$

As pointed out by \citet{jax2018github,jeon2021differentiable}, this can be applied more generally. Start with some contraction mapping $x = f(\theta, x)$ which implies a solution function $\theta \mapsto x^*(\theta)$, where we want to retrieve the reverse-mode derivative (vector-Jacobian product) $v' \mapsto v'\partial x^*(\theta)$. Assume that the implicit function theorem holds, and that $x^*(\theta) = f(\theta,x^*(\theta))\ \forall\ \theta$ in some neighborhood around $\theta_0$ which we want to differentiate. Differentiating both sides of the equation, we have

$$ \partial x^*(\theta_0) = B + A\partial x^*(\theta_0) = (I-A)^{-1}B $$

where $A = \frac{\partial f(\theta_0, x^*(\theta_0))}{\partial x^*(\theta_0)},\ B = \frac{\partial f(\theta_0, x^*(\theta_0))}{\partial \theta_0}$. Therefore, we can evaluate vector-Jacobian products by

$$ v'\partial x^*(\theta_0) = v'(I-A)^{-1}B = w'B $$

where $w' = v'(I-A)^{-1} = v' + w'A$. We can equivalently think of $w'$ as the fixed point of the map $u' \mapsto v' + u'A$, so we can write the reverse mode derivative of a contraction mapping as the solution to another contraction mapping, avoiding the expensive operation of matrix inversion.

\section{Algorithmic Improvements}\label{sec:algo_improv}

\subsection{Optimization Methods}

\paragraph{Quasi-Newton Methods}\mbox{}

Most of the BLP literature use quasi-Newton methods for frequentist optimization and estimation. When the objective function is convex, quasi-Newton methods that utilize an approximation to second-order Hessian information such as BFGS and its variants will dominate a gradient-descent based approach that does not. However, in the case of highly non-convex objective functions that can be ill-behaved outside a feasible region such as the BLP objective function, quasi-Newton methods can have issues with numerical stability and fail to converge \citet{knittel2014estimation}. \citet{conlon2020best} resolve this by noting that imposing parameter bounds helps with numerical stability, both in replications of \citet{berry1995automobile,nevo2001measuring}, as well as simulation exercises. We are able to confirm this finding. However, the choice of parameter bounds is a subjective hyper-parameter that can be difficult to justify in empirical work, motivating the search for an optimization method that functions well even without parameter bounds.

\paragraph{Modern Gradient Descent Methods}\mbox{}

For ill-conditioned problems, gradient descent methods which only utilize first-order derivative information controlled by the learning rate offers more consistent results without parameter bounds across a range of reasonable starting values. The challenge here is to have an optimization method that is relatively robust to local stationary points, but still achieves good convergence. There are two main branches of modern gradient descent algorithms, Stochastic Gradient Descent (SGD) and adaptive learning rate or ``momentum'' based algorithms \citep{dogo2018comparative}.

In SGD (or mini-batch SGD mSGD) \citep{schaul2013unit}, each update is computed using the gradient with respect to a single (small batch of) observation(s). This helps with local stationary points because the stationary point of the objective function with respect to each (batch of) observation(s) is likely to be different, and this difference will allow the optimizer to ``escape'' local minima. We do not test or employ SGD in our implementation because it is not amenable to Just-In-Time-compilation (JIT-compilation). The only principled way to batch the data within the BLP setting is to group them by market. However, doing so will change the dimension of the design matrices at each iteration of the objective function, defeating JIT-compilation. SGD also relies on annealing the learning rate over iterations to control the over-shoot from using batched gradient descent, which is hard to calibrate in practice.

Adaptive learning rate optimization algorithms such as RMSProp \citep{hinton2012neural}, AdaGrad \citep{duchi2011adaptive}, and Adam \citep{kingma2014adam,reddi2019convergence} adjust the learning rate on a per-parameter basis based on a memory of past gradients. The intuition here is that if there is a sudden and large change in the direction of the optimization update, then we should be cautious about moving in that direction. Hence, momentum-based optimizers keep a ``memory'' of previous gradients, and generates an update that attenuates sudden, large changes and amplifies movements in consistent directions. This momentum simultaneously addresses the exploding gradient problem and allows the optimizer to escape local optima. We use the AdaBelief algorithm \citep{zhuang2020adabelief} implemented in \texttt{Optax} \citep{optax2020github}. It is an Adam variant which incorporates curvature information without approximating the Hessian by utilizing deviations between predicted and actual gradients. We find that with adaptive learning rate gradient descent optimizers, imposing parameter bounds is no longer necessary, removing another source of hyper-parameter arbitrariness in estimation.

\subsection{MCMC Sampling}
\paragraph{Choice of Sampler}\mbox{}

\citet{chernozhukov2003mcmc} use a RWMH in their original paper on LTEs. However, \citet{forneron2020inference} find that RWMH takes a long time to converge for the BLP problem, resulting in a low effective sample size even with a long Markov chain. \citet{hong2020blp} propose using Hamiltonian Monte Carlo (HMC) \citep{duane1987hybrid,neal2011mcmc} instead, which utilizes gradient information in sampling.\footnote{\citet{betancourt2017conceptual} provides a comprehensive introduction to HMC and its variants.} We can think of the proposal process of RWMH as a blind-folded person throwing a ball with random direction and momentum. Wherever the ball lands, we check the value of the objective function relative to our current position, and assign a high acceptance probability if there is a large decrease in the objective function value. In HMC, we can think of the ground as having gravity wells, where regions of low objective function value have a higher gravity than regions of high objective function value. The same ball with random direction and momentum no longer travels in a straight line, but is instead pulled towards regions of low objective function value, speeding convergence. Because the ball is thrown with random momentum, there is still a chance that the sampler will escape the pull of the gravity wells and therefore explore beyond the closest local minimum. The curved path of the ball represents an integral that is hard to compute exactly. In practice, a simplectic integrator based on Hamiltonian dynamics is used, breaking up the curved path into $L$ steps of length $\varepsilon$, which are the hyper-parameters of the sampler.

We extend the work of \citet{hong2020blp} by using a variant of Hamiltonian Monte Carlo that is less sensitive to the choice of hyper-parameters. HMC's hyper-parameters are $(L, \epsilon, M)$, which are the number of steps in the leap-frog symplectic integrator, the size of each step in the integrator, and the mass matrix or the covariance matrix of the proposal distribution respectively\footnote{Correspondence with \citet{hong2020blp} authors confirm their choice of parameters as $L=10,\ \epsilon=0.1,\ M = I$, as in \citet{neal2011mcmc}.}. The tuning of these hyper-parameters is artisanal and problem-specific, and \citet{hoffman2014no} outline the effects of a poorly chosen set of $(L, \epsilon, M)$. An $L$ that is too large results in wasted computation, and in the converse results in a poor exploration of the parameter space. $\epsilon$ controls the granularity of exploration, and a well-chosen $M$ that approximates the posterior covariance matrix increases the rate of convergence of the Markov Chain. They propose a No-U-Turn Sampler (NUTS) variant of HMC that automatically tunes the hyper-parameters. Instead of a single ball thrown in one direction, NUTS uses two balls thrown in opposite directions with the same random momentum. It grows the trajectory of both balls using the simplectic integrator (choosing which to advance at random) until one of the paths turns on itself, indicating that remainder of the path is likely to backtrack. It then samples a point along this bi-directional path as the proposed candidate. This is an auto-tuning of the number of steps parameter $L$ on a per-sample basis. $\epsilon$ and $M$ are tuned in a set of warm-up draws using prime-dual averaging for a target acceptance rate, and the empirical covariance matrix of the warm-up draws respectively. With minor modifications, NUTS-HMC has become the standard in the machine learning literature \citep{carpenter2017stan, phan2019composable,salvatier2016pymc3}, and thus our MCMC sampler of choice as implemented in \texttt{NumPyro} \citep{phan2019composable}.

\paragraph{Sampler Settings and Diagnostics}\mbox{}

Rather than a single long chain of MCMC samples that is common in the economics literature \citep{hong2020blp,chernozhukov2003mcmc,forneron2020inference}, we use many smaller chains so that we can compute MCMC mixing diagnostics that are common in the computer science literature. The use of a single Markov chain can lead to a failure to diagnose non-convergence, and this has been noted as far back as \citet{gelman1992single}. In particular, the seeming convergence of a Markov chain to a single mode may fail to detect multi-modality that is far apart -- which in economics implies a potential identification failure. Similarly, the performance of a single chain may fail to detect that the chain is stuck in a region of high-curvature, with a step-size that is too large to allow for acceptable proposals to ``escape'' -- which can imply the brittleness of frequentist estimation through optimization. We use the rank-normalized split-$\hat{R}$ of \citet{vehtari2021rank} as a diagnostic, which checks for whether the variance of the first half of a chain fails to match the second half, and whether the variance of multiple chains together is greater than each individual chain. An affirmative of either condition shows up as a high $\hat{R}$, and indicates a failure to converge. The $\hat{R}$ is an oft-recommended convergence diagnostic used across a range of MCMC statistical packages \citep{salvatier2016pymc3,carpenter2017stan}. \citet{carpenter2017stan} suggests using at least four Markov chains in estimation using a split-$\hat{R}$ cut-off of 1.05 to indicate a lack of convergence.

MCMC sampling can also produce draws which are auto-correlated, which reduces the effective region explored by the sampler. This is clear by construction with a Gibbs sampler, but persists even with RWMH. Therefore, one common diagnostic is to compute the effective sample size $N_{eff}$ of the sampler, accounting for this auto-correlation \citep{forneron2020inference}. If $N_{eff} \ll N$, where $N$ is the number of samples to be drawn, then this could indicate numerical issues with the MCMC sampler.\footnote{HMC-NUTS attempts to perform anti-correlated draws, so in fact $N_{eff} > N$ is possible.} We use the rank-normalized version of $N_{eff}$ \citep{vehtari2021rank}, which abstracts away potential issues with heavy-tailed distributions and reparametrization.

\section{Empirical Exercises}\label{sec:mc_setup}

Across all empirical exericse, we use the $R_d$ sequence of \citet{roberts2018unreasonable} for numerical integration. It has good performance while eschewing the need to tune any hyper-parameters such as co-primes for Halton sequences. All empirical exercises were run GPU-accelerated on consumer hardware, inside the Docker container provided by \texttt{admest}.\footnote{The actual system comprises an Intel i9-10850K CPU and an Nvidia RTX 2070 Super GPU running Fedora 34 as the host operating system. The Docker container builds on an Ubuntu 18.04 image provided by Nvidia for CUDA/CUDNN development.} For the 2-Stage GMM models, we use the weight matrix

$$ W = \left[Z'\Omega_{\xi}Z\right]^{-1} $$

instead of the CUE weight matrix specified in Section \ref{sec:blpdo} to prevent numerical issues when computing the analytical standard errors.

\subsection{Monte Carlo Simulations}

Our simulation set-up largely mirrors that of \citet{conlon2020best}. We randomly draw utilities and costs, but solve for equilibrium prices and shares. Over 100 different synthetic datasets, we set $T=20$ markets and choose a random number of firms from 2 to 10, and then have each firm produce between 3 to 5 products. Similar to \citet{conlon2020best}, this procedure yields $200 < N < 600$ on average.

In order to make the choice of weight matrix consequential, we introduce heteroskedasticity by construction into the structural erorrs $(\xi_{jt},\omega_{jt})$. we first draw $(\psi^1_{jt}, \psi^2_{jt})$ from independent normal distributions with variance 0.2, and $\phi_{jt}^1,\ \phi_{jt}^2 \sim Unif(0.5, 2)$. We then taking $\xi_{jt} = \psi^1_{jt} \times \phi_{jt}^1,\ \omega_{jt} = \psi^2_{jt} \times \phi_{jt}^2$ as the structural errors for the simulation.  Linear demand characteristics and supply characteristics follow the same formulation as \citet{conlon2020best} with $[1,x_{jt},p_{jt}]$ and $[1,x_{jt},w_{jt}]$ respectively. The two exogenous characteristics are drawn from the standard uniform distribution, and we assume that the coefficient on $x_{jt}$ is random. We assume Bertrand competition for our conduct matrix, and a linear function on marginal costs so $c_{jt} = [1,x_{jt},w_{jt}]\theta_3 + \omega_{jt}$. The endogenous prices and shares are computed using the $\zeta$-markup approach of \citet{morrow2011fixed}. In simulating the data, we use 1,000 draws from the $R_d$ sequence to perform numerical integration, to prevent fluctuations across simulations arising from integration error. We use similar true values to \citet{conlon2020best}, with $\theta_1 = [-7,6,1],\ \theta_2 = [3],\ \theta_3 = [2,1,0.5]$. This generates $Corr(p_{jt},w_{jt}) \approx 0.2$, implying relatively weak cost-shifting instruments, and outside shares that average around 0.9.

We estimate both Demand-Only and Demand-Supply BLP models, drawing starting values from a uniform distribution (common to each simulation run) with support 50\% above and below the true value. We use 100 $R_d$ sequence draws for numerical integration within the contraction mapping. For Demand-Only BLP, we estimate four models, given by

\begin{enumerate}[a)]
    \item 2S-GMM with analytical gradients (AG 2S) similar to \citet{conlon2020best}, using L-BFGS-B\footnote{We use \texttt{NLopt}'s implementation rather than \texttt{SciPy} since it natively supports sharing calculations between value and gradient computations.}, box constraints from 10\% to 1,000\% below and above the true value, and a $L^\infty$ tolerance on objective function improvements of \texttt{1E-10}.
    \item 2S-GMM with automatic differentiation (AD 2S) and the AdaBelief optimizer configured with a learning rate of 0.1, an $L^\infty$ tolerance on the gradient of \texttt{1E-10}, and no bounds.
    \item CUE with same settings as b).
    \item LTE estimation using NUTS-HMC, targetting the standard acceptance rate of 0.8, and four chains of 500 samples, with a discarded warm-up period that is 20\% of the number of final samples (that is, 100).
\end{enumerate}

For Demand-Supply BLP, we estimate models b) and c) under similar settings. For the LTE estimator, we report uncertainty quantification based on the credible intervals of the sample in addition to computing the frequentist standard errors from a point estimate such as the posterior mean.

\section{Results}\label{sec:mc_results}
\paragraph{Relative Performance of Different Models}\mbox{}\\
For each model, we report the following statistics across all demand-related parameters (including all linear parameters, which are often of interest also): the mean bias, the Median Absolute Error (MAE), as well as the empirical coverage.

\begin{table}[!htbp] \centering 
\caption{Monte Carlo Simulation Results (50 Simulation Runs)} 
\label{tab:simresults}
\makebox[\textwidth]{
\begin{threeparttable}
\begin{tabular}{cccccccccc} 
\\[-1.8ex]\hline 
\hline \\[-1.8ex] 
& & \multicolumn{4}{c}{Average Bias} & \multicolumn{4}{c}{Empirical Coverage} \\
Variant & Model & $\theta_{1,int}$ & $\theta_{1,x}$ & $\theta_{1,price}$ & $\theta_{2,x}$ & $\theta_{1,int}$ & $\theta_{1,x}$ & $\theta_{1,price}$ & $\theta_{2,x}$ \\
\hline \\[-1.8ex] 

\multirow{6}{*}{Demand-Only}
& AG 2S & -0.142 & -0.18 & 0.049 & 0.14 & 0.96 & 0.9 & 0.96 & 0.86 \\
& & (0.608) & (0.488) & (0.179) & (0.435) \\
& AD 2S & -0.178 & -0.191 & 0.06 & 0.147 & 0.96 & 0.92 & 0.96 & 0.86 \\
& & (0.579) & (0.417) & (0.197) & (0.429) \\
& CUE & -0.169 & -0.181 & 0.057 & 0.131 & 0.94 & 0.94 & 0.94 & 0.92 \\
& & (0.677) & (0.503) & (0.198) & (0.415) \\
& LTE$^{\dagger}$ & -0.21 & -0.173 & 0.068 & 0.1 & 0.2 & 0.98 & 0.4 & 0.92 \\
& & (0.709) & (0.513) & (0.207) & (0.432) & (0.94) & (0.92) & (0.94) & (0.88) \\
\hline \\[-1.8ex] 

\multirow{4}{*}{Demand-Supply}
& AD 2S & 0.578 & -0.111 & -0.168 & 0.325 & 0.92 & 0.94 & 0.94 & 0.84 \\
& & (0.697) & (0.419) & (0.207) & (0.426) \\
& CUE & -0.21 & -0.069 & 0.064 & 0.027 & 0.94 & 0.98 & 0.94 & 0.96 \\
& & (0.705) & (0.443) & (0.205) & (0.348) \\
\hline \\[-1.8ex] 
\end{tabular}
\begin{tablenotes}
\setlength\labelsep{0pt}
\item \textit{Median Absolute Error (MAE) reported in parantheses under average bias.}
\item \textit{LTE results excludes divergent chains and chains where $\hat{R} >  1.05$.}
\item \textit{$^{\dagger}$Frequentist uncertainty quantification (in paratheses below credible intervals coverage) should be used for concentrated out parameters in LTE estimation.}
\end{tablenotes}
\end{threeparttable}}
\end{table}

Our simulation results for the CUE confirm previous theoretical results from \citet{hansen1996finite,newey2009generalized,newey2004higher,donald2000jackknife}. The CUE and LTE (which uses the CUE objective function) estimators have a lower bias but higher MAE (recall that $Corr(p_{jt},w_{jt}) \approx 0.2$ implies relatively weak instruments) compared to 2S-GMM. This is especially true with the inclusion of a supply-side. Of particular note is the substantially lower bias for $\theta_{1,price},\ \theta_{2,x}$ for the CUE estimator over the 2S-GMM estimator, which are often the key parameters of interest. This could be due not just to the inherently lower bias of the CUE, but also because the CUE formulation allows for the linear parameter on price $\alpha = \theta_{1,price}$ to be done within the objective function itself, which forces it into a feasible region.

Similar to \citet{hong2020blp}, we find that the 2S-GMM estimators can have empirical coverage that is different from the nominal level. Our results differ in that we find under-coverage of the non-linear parameter $\theta_{2,x}$ rather than over-coverage. Similar to \citet{hong2020blp}, using the LTE brings the empirical coverage close to the nominal level, but we find this using the credible interval rather than using the LTE as a point-estimate for frequentist uncertainty quantification. We also find that the CUE estimator produces similar results or slightly better results in terms of coverage, which suggests that the improvements in coverage could be due to the use of the CUE objective function rather than the use of LTE estimation. We find that using credible intervals from the sampler directly for non-linear parameters alongside frequentist uncertainty quantification for the linear parameters delivers the best results for LTE uncertainty quantification.

\paragraph{MCMC Diagnostics}\mbox{}\\
For each varaint of the simulation exercise, we report the average $\hat{R}$, the number of runs with $\hat{R} > 1.05$, the average $N_{eff}$, and the number of divergent chains.

\begin{table}[htbp!]
\centering
\caption{MCMC Diagnostics for Simulation Exercise}
\label{tab:mcmc_diagsim}
\begin{threeparttable}
\begin{tabular}{cccc}
\\[-1.8ex]\hline 
\hline \\[-1.8ex] 
Variant & Average $\hat{R}$ & Average $N_{eff}$ & Divergent Chains \\
\hline \\[-1.8ex] 
\multirow{2}{*}{DO Simple} & 1.007 & 670.253 & 0\% \\
& (0\%) \\
\hline \\[-1.8ex] 
\end{tabular}
\begin{tablenotes}
\setlength\labelsep{0pt}
\item \textit{Percentage of runs with $\hat{R}$ above 1.05 in parantheses.}
\end{tablenotes}
\end{threeparttable}
\end{table}

Contrary to \citet{forneron2020inference}, we find that the LTE estimated using NUTS-HMC when applied to the BLP problem has good MCMC diagnostics. Our $N_{eff}$ is high despite modest chain sizes, and the average $\hat{R}$ indicates a lack of evidence that our Markov chains fail to mix. We also have little to no divergent chains. The difference is likely due to our use of HMC-NUTS rather than RWMH for sampling.

\section{Conclusion}\label{sec:conclude}

We have shown how to phrase the BLP problem as an automatically differentiable moment function that is numerically equivalent to implementations in the literature based on for-loops. The use of automatic differentiaton allows us to implement more theoretically appealing estimators such as the CUE. Within the BLP setting, it also allows us to compute the linear parameter on price $\alpha$ ``online'' within the objective function itself since the weight matrix is no longer shared and we do not need to use multiple-equation GMM. This has resulted in better performance without the use of parameter bounds since $\alpha$ is forced into a feasible region. We anticipate that these techniques should be broadly applicable to other econometric models, particularly in situations where we have concentrated out parameters or sophisticated structural models.

The use of automatic differentiation has also enabled us to draw on the latest in machine learning estimation techniques. We provide intuition for why adaptive learning rate optimization methods are better suited to the BLP problem, and how NUTS-HMC sampling is superior to traditional RWMH for LTE estimation. Since automatically differentiable modifications preserve automatic differentiability, we envision that the robust estimation of BLP extensions and variants will now be much easier. For example, one can assume a dense covariance matrix on the random coefficients as opposed to a diagonal matrix as in the original formulation, or extend BLP to dynamic models of demand. 

The implementation in \texttt{admest} outlined by this paper brings algorithmic improvements that better leverage modern hardware such as GPUs and TPUs for econometric estimation. We also incorporate hitherto neglected best practices for MCMC sampling that are applicable for all LTE-type estimation, and quasi- or full Bayesian approaches to econometric estimation more generally. The inclusion of adapative learning rate first order gradient descent methods should help with numerical stability for poorly behaved objective functions where traditional quasi-Newton methods tend to diverge. Once again, these advances should prove useful in broad classes of econometric estimation problems beyond BLP.

Our simulation exercises extend the work of \citet{conlon2020best,hong2020blp}, and allow us to examine past theoretical work on the CUE and many weak moment conditions more generally \citep{hansen1996finite,newey2009generalized,newey2004higher,donald2000jackknife} within the context of a non-linear IV-GMM with concentrated out parameters. Simulation results affirm that CUE-based frequentist and LTE estimation has a lower bias but higher MAE compared to traditional 2S-GMM methods. We also find that asymptotic normality is likely a poor approximation to the ``medium sample'' distribution of the BLP estimator, motivating the choice of the LTE over traditional frequentist uncertainty quantification, at least for the non-linear parameters that are directly sampled.

Future work can incorporate the bias correction of \citet{newey2004higher,chen2019mastering}, which will not only improve the bias of the BLP estimator, but will also likely lead to empirical coverage closer to the nominal level. With automatic differentiation, the implementation of analytical corrections should be straightforward, since derivatives are ``free''; and with advances in computation the time-costs of implementing empirical corrections should also decrease. The framework of automatically differentiable M-estimation can also be applied to structural estimation more generally, both in the field of industrial organization and beyond.

\newpage
\printbibliography

\newpage
\appendix
\section{Demand-Only 2S-GMM NFP}\label{app:do2s}
We present the 2S-GMM NFP objective function, gradient, and uncertainty quantification, deriving the relevant analytical gradients.

\subsection{Objective Function}
\begin{algorithm}
\DontPrintSemicolon
\KwData{$X_1 \in \mb{R}^{J\times K_1},\ X_2 \in \mb{R}^{J\times K_2}$ the design matrices of features that enter linearly and non-linearly respectively; $Z \in \mb{R}^{J \times z}$ the matrix of instruments; $\nu \in \mb{R}^{R \times K_2}$ the matrix of simulation draws; $\mc{S} \in \mb{R}^J$ the vector of observed shares; $\iota_{J/J}, \iota_J$ as in Section \ref{sec:blpdo_obj}.}
\KwIn{The value of $\theta_2$ being considered, the starting point for the fixed point equation $\delta_0$, the weighting matrix $W$.}
\KwResult{The GMM objective function $q(\theta_2)$}
\Begin{
    $\mb{R}^{J\times R} \ni \mu(\theta_2) \longleftarrow X_2 \Omega_{\theta_2} \nu'$\;
    $\mb{R}^J \ni \hat{\delta} \longleftarrow \lim_{n\to\infty}\delta_n(\delta_0; \mu(\theta_2))$\;
    $\mb{R}^{K_1} \ni \hat{\theta_1} \longleftarrow \left[X_1'ZWZ'X_1\right]^{-1}X_1'ZWZ'\hat{\delta}$\;
    $\mb{R}^J \ni \xi \longleftarrow \hat{\delta} - X_1\hat{\theta}_1$\;\;

    $q(\theta_2) \longleftarrow \xi'Z W Z'\xi$
}
\caption{NFP GMM Objective}
\label{demand_obj_algo}
\end{algorithm}

The (protected) fixed-point equation is computed using

$$ \mb{R}^{J \times R} \ni \hat{s}(\delta_n;\mu(\theta_2)) = \exp(\delta_n +^c \mu(\theta_2) - \iota_{J/J} [\delta_n +^c \mu(\theta_2)]) \oslash \left[1 + \iota_J\exp(\delta_n +^c \mu(\theta_2) - \iota_{J/J} [\delta_n +^c \mu(\theta_2)])\right] $$
$$ \delta_{n+1}(\delta_n; \mu(\theta_2)) = \delta_n + \log(S) - \log(rowSum(\hat{s}(\delta_n;\mu(\theta_2)))) $$

To compute the estimate for the concentrated out parameters $\hat{\theta_1}$, we follow \citet{nevo2000practitioner} and \citet{conlon2020best} in using a single-stage linear GMM with the weighting matrix of the outer (non-linear) GMM loop $W$. This approach greatly simplifies the computation of the analytic standard errors.

\subsection{Gradient}\label{demand_grad}
The gradient of the GMM objective function is given by

$$ \nabla q(\theta_2) = 2 \frac{\partial\delta}{\partial\theta_2}'Z W Z'\xi $$

\noindent{}Following \citet{nevo2000practitioner} in invoking the implicit function theorem, at the market level we have

$$ \frac{\partial\delta}{\partial\theta_2} = -\left[\frac{\partial \hat{s}}{\partial\delta}\right]^{-1} \frac{\partial \hat{s}}{\partial\theta_2} $$

\begin{align*}
\frac{\partial \hat{s}_j}{\partial\delta_m} 
&= \begin{cases}
\int \hat{s}_{ij}(1-\hat{s}_{ji})d(\nu)\ \forall\ j=m \\
\int \hat{s}_{ji}\hat{s}_{mi}d(\nu)\ \forall\ j\neq m
\end{cases} \\
\frac{\partial \hat{s}_j(\theta_2, \delta)}{\partial\theta_{2m}}
&= \int \nu_{ik}s_{ji}\left(x_{jk} - \sum_{m=1}^Jx_{mk}s_{mi}\right)d(\nu)
\end{align*}

\noindent{}For $\frac{\partial \hat{s}}{\partial\delta}$, we can use $\hat{s}\hat{s}'$ for the off-diagonal elements, and for the diagonal the row means of $\hat{s} \odot (1-\hat{s})$ where the subtraction is element-wise. For the second matrix, under the assumption of overlapping draws and pseudo- or quasi-Monte Carlo integration nodes, we have the expression

$$ \frac{\partial \hat{s}}{\partial\theta_2} = \left(\hat{s}\nu\right)\odot X_2 - \hat{s}\left(\nu \odot [\hat{s}'X_2]\right) $$

\noindent{}Now we show how to move from the single market scenario to computing all markets without any loops. Intuition might suggest that we can simply substitute the market level matrices for the overall matrices with all markets. This approach is on the right track, but requires some modifications. Across all markets, the matrix $\frac{\partial \hat{s}}{\partial\delta}$ is given by the block diagonal matrix

$$ \frac{\partial \hat{s}}{\partial\delta} = 
\begin{bmatrix}
    \frac{\partial \hat{s}}{\partial\delta}_1 & 0           & \cdots & 0 \\ 
    0           & \frac{\partial \hat{s}}{\partial\delta}_2 & \cdots & 0 \\
    0           & 0           & \ddots & 0 \\
    0           & 0           & \cdots & \frac{\partial \hat{s}}{\partial\delta}_T
\end{bmatrix} $$

\noindent{}where $\frac{\partial \hat{s}}{\partial\delta}_t$ is the market level matrix for market $t$. Substituting the market level share matrix with the overall matrix will yield the correct entries in the block diagonal, but also include non-zero values in the off block diagonal entries. Therefore, we use $\iota_J$ as a template to ``slice'' out the entries we want through Hadamard (element-wise) multiplication. Ignoring the diagonal entries (for which we can substitute the market level matrix with the overall matrix), we have

$$ \frac{\partial \hat{s}}{\partial\delta} = \hat{s}\hat{s}' \odot \iota_J $$

\noindent{}It is clear from this construction that computing $\frac{\partial \hat{s}}{\partial\delta}$ in this way requires more operations than looping over markets since the latter avoids computing the off block diagonal entries. The degree of inefficiency will depend on the sparsity of the matrix $\frac{\partial \hat{s}}{\partial\delta}$, which in turn depends on the uniformity of the number of products in each market. If the number of products in each market is similar, then there will be more unused off block diagonal entries computed.

\begin{figure}[htbp!]
    \centering
    \begin{subfigure}{.3\columnwidth}
        \centering
        \includegraphics[width=\columnwidth]{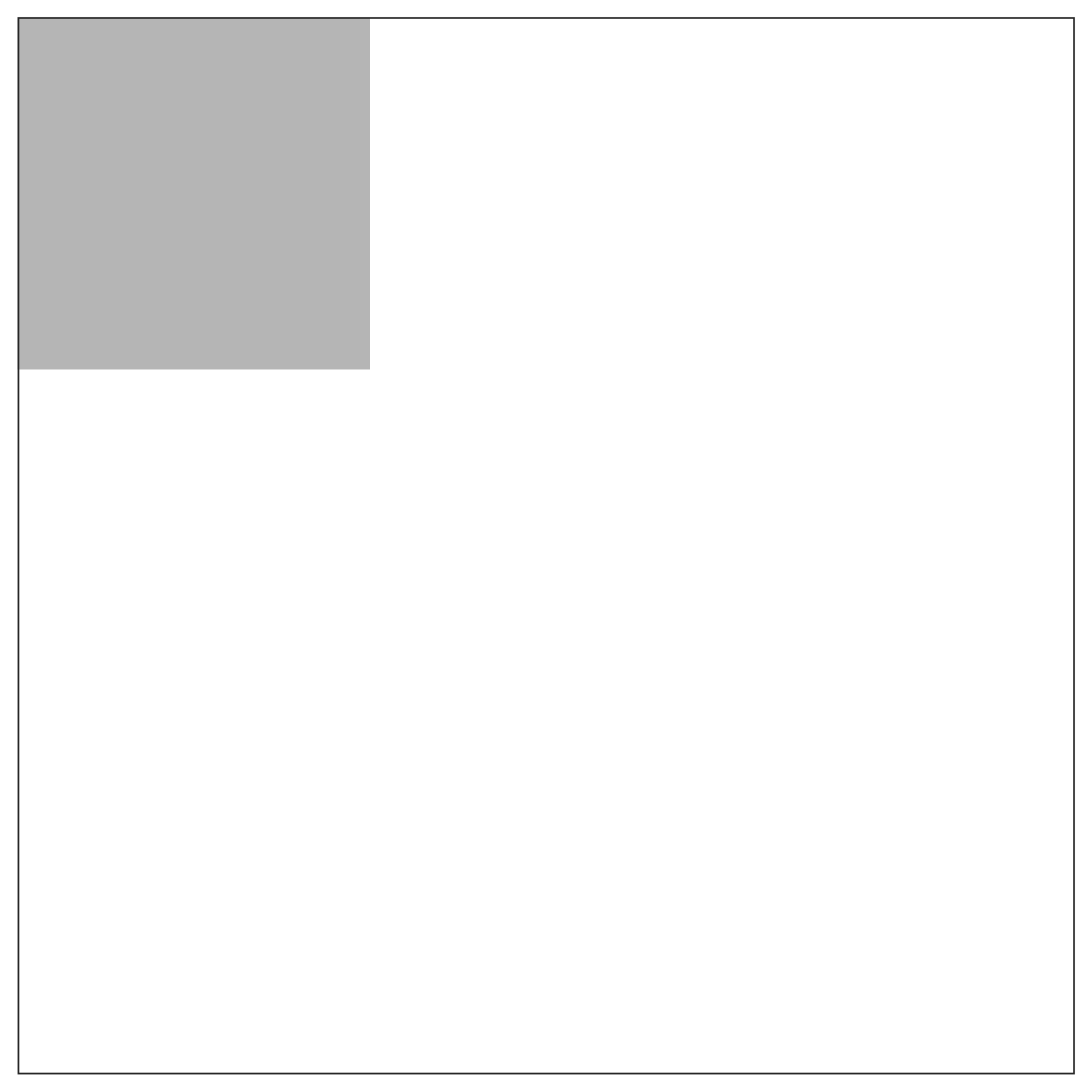}
    \end{subfigure}
    \begin{subfigure}{.3\columnwidth}
        \centering
        \includegraphics[width=\columnwidth]{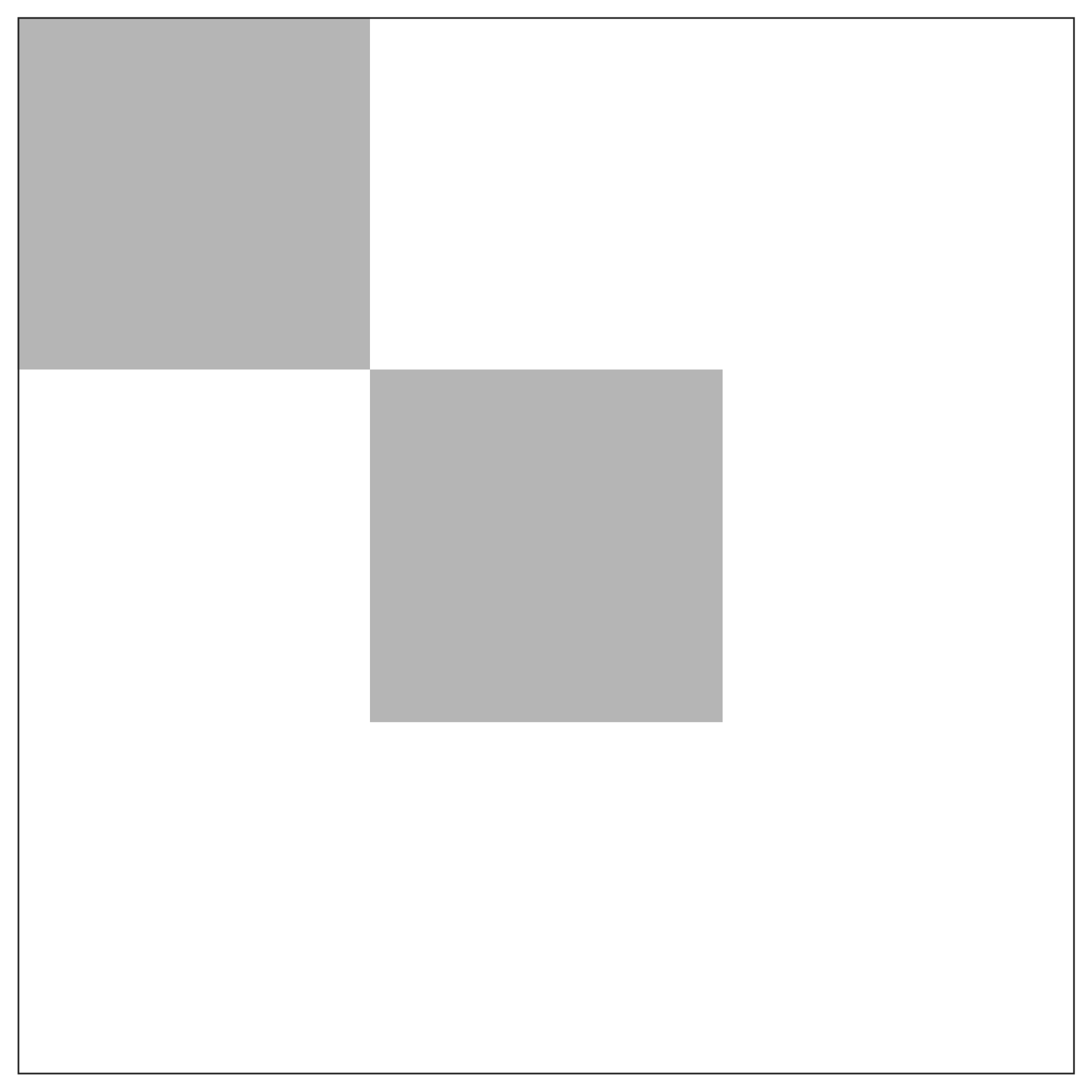}
    \end{subfigure}
    \begin{subfigure}{.3\columnwidth}
        \centering
        \includegraphics[width=\columnwidth]{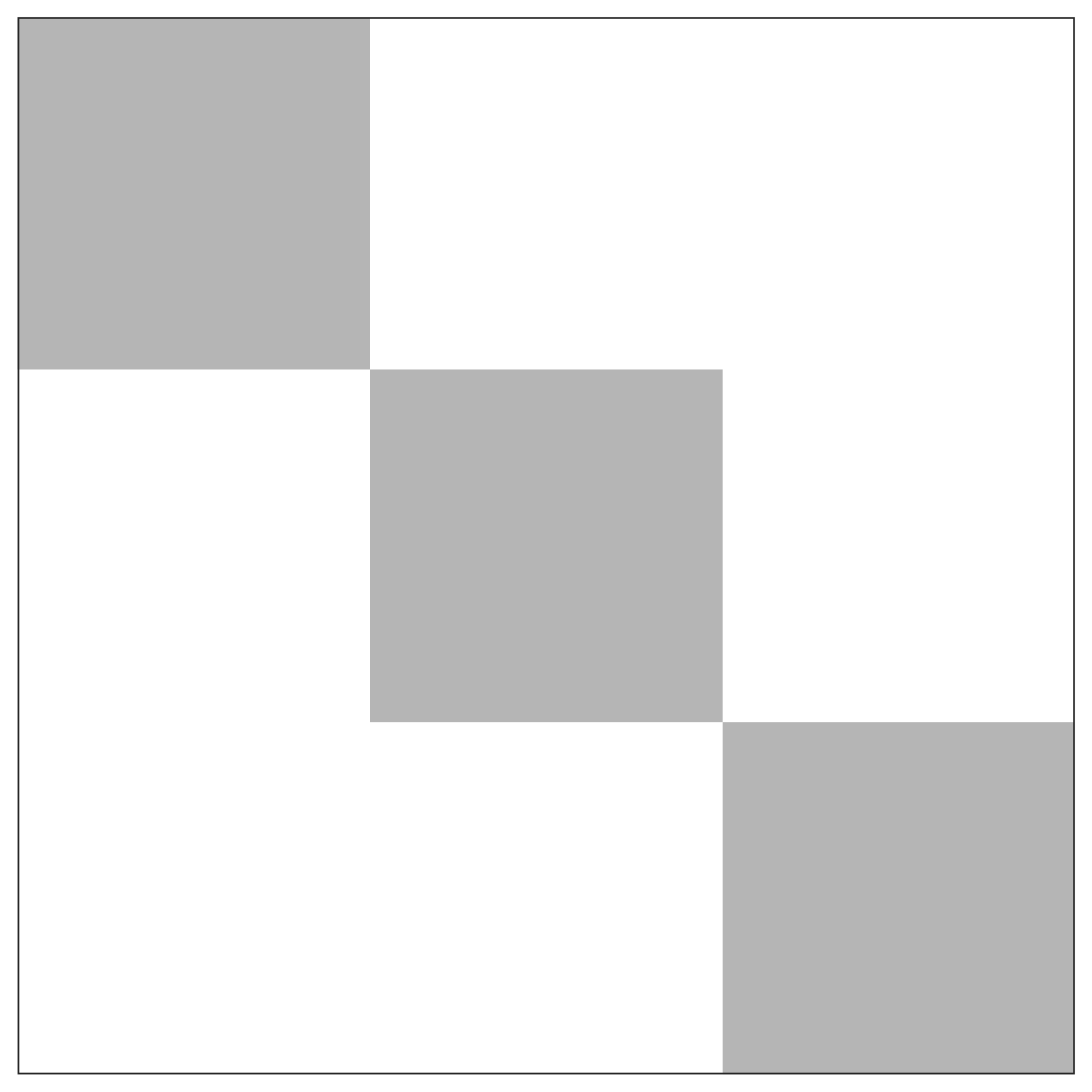}
    \end{subfigure}
    \caption{Building $\frac{\partial \hat{s}}{\partial\delta}$ using a Loop}
\end{figure}

\begin{figure}[htbp!]
    \centering
    \begin{subfigure}{.3\columnwidth}
        \centering
        \includegraphics[width=\columnwidth]{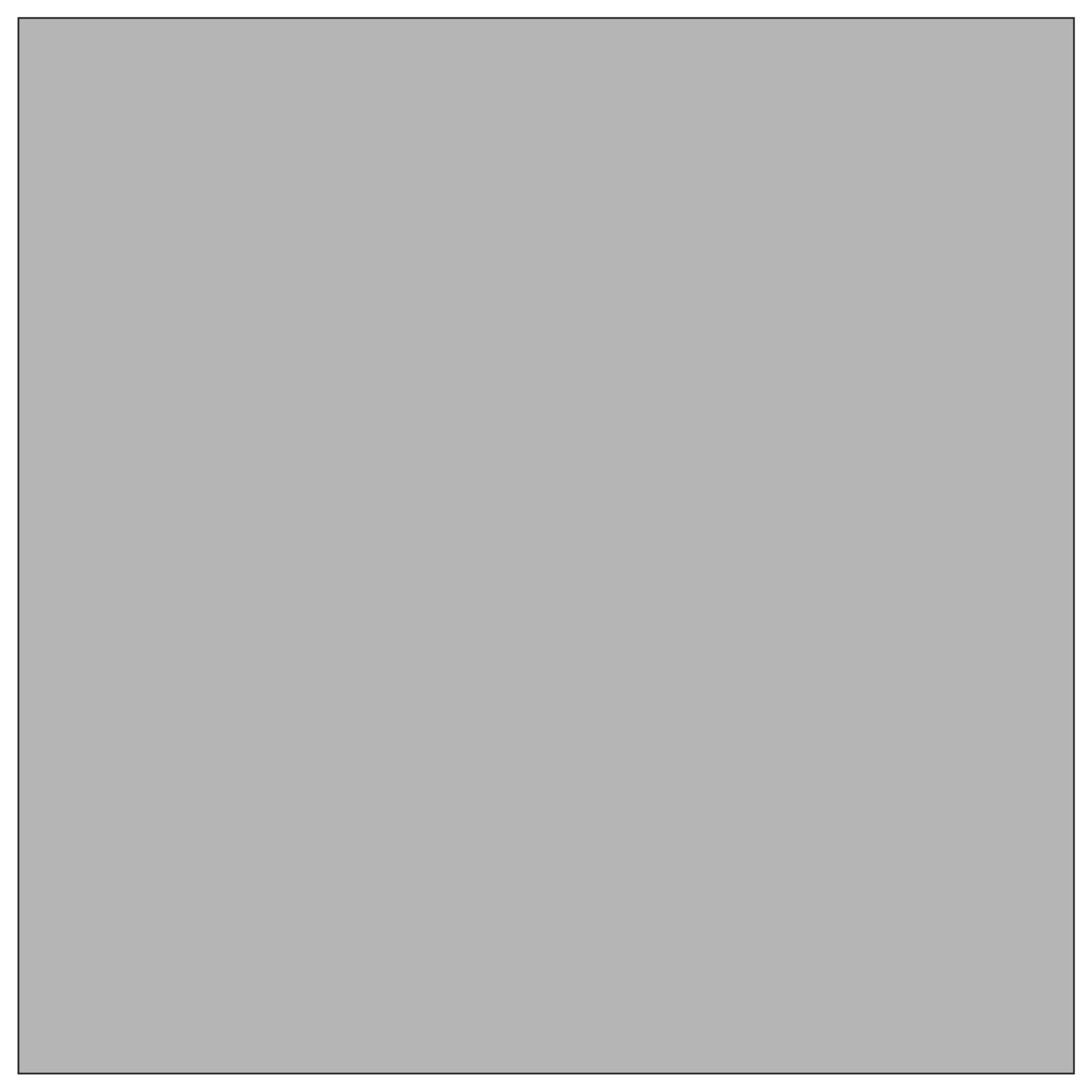}
    \end{subfigure}
    \begin{subfigure}{.3\columnwidth}
        \centering
        \includegraphics[width=\columnwidth]{zz_fig_loop3.png}
    \end{subfigure}
    \caption{Building $\frac{\partial \hat{s}}{\partial\delta}$ using a $\iota_J$}
\end{figure}

\begin{figure}[htbp!]
    \centering
    \begin{subfigure}{.3\columnwidth}
        \centering
        \includegraphics[width=\columnwidth]{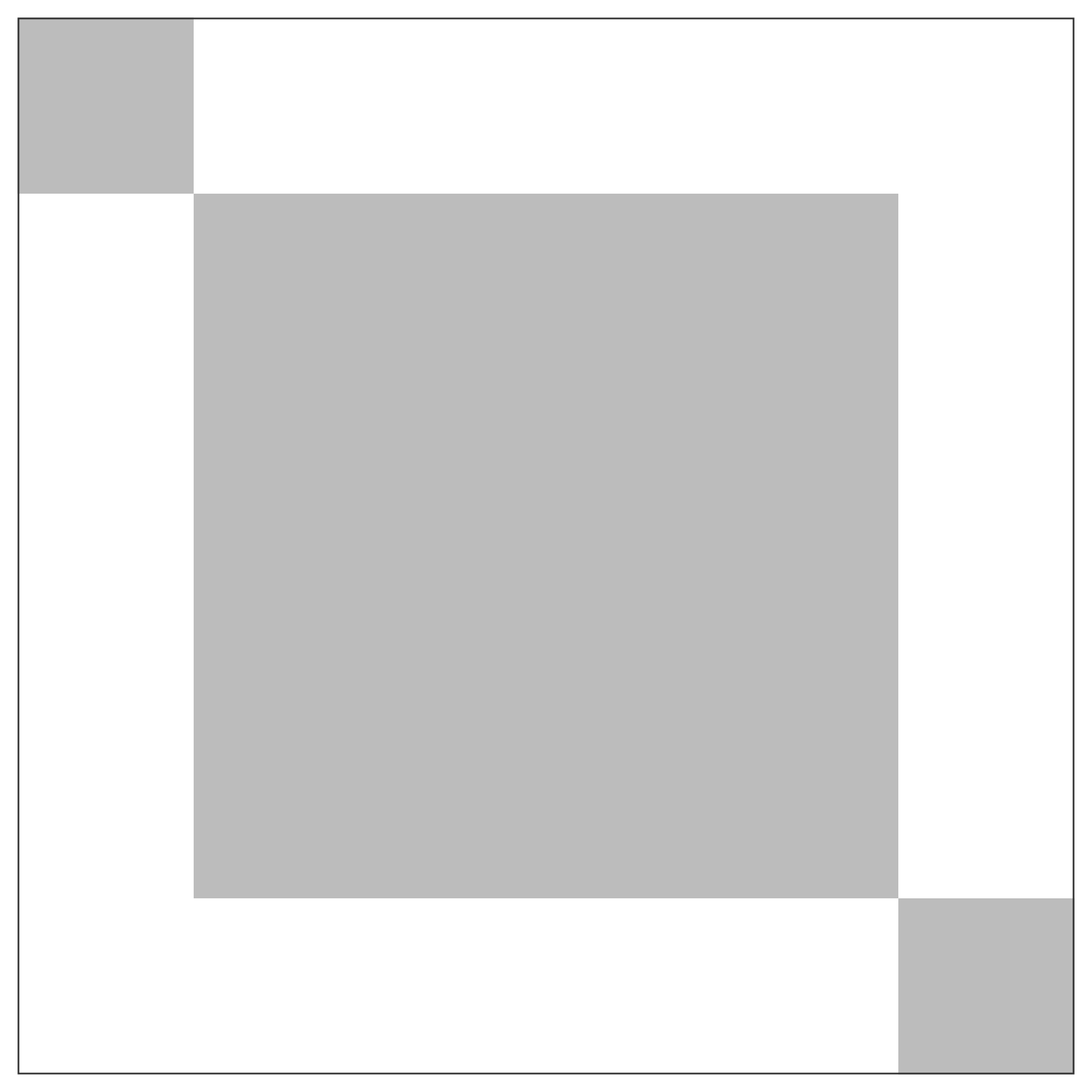}
        \caption{Less Extra Computation}
    \end{subfigure}
    \begin{subfigure}{.3\columnwidth}
        \centering
        \includegraphics[width=\columnwidth]{zz_fig_loop3.png}
        \caption{More Extra Computation}
    \end{subfigure}
    \caption{Extra Computation Determined by Evenness of Number of Products}
\end{figure}

\newpage
Now we turn our attention to $\frac{\partial \hat{s}}{\partial\theta_2}$. The first term can be computed by substituting the market level matrices with the overall matrices. However, the second term $\hat{s}\left(\nu \odot [\hat{s}'X_2]\right)$ cannot be. For the moment, ignore the Hadamard multiplication by $\nu$. Then what we want is a stacked column of block matrices

$$ \begin{bmatrix}
(\hat{s}\hat{s}'X_2)_1 \\ \vdots \\ (\hat{s}\hat{s}'X_2)_T
\end{bmatrix} $$

\noindent{}but if we simply substitute the overall matrices, this will yield

$$ \begin{bmatrix}
\hat{s}_1\left[\sum_t (\hat{s}'X_2)_t\right] \\
\vdots \\
\hat{s}_T\left[\sum_t (\hat{s}'X_2)_t\right]
\end{bmatrix} $$

\noindent{}Instead, define $\hat{s}_D$ to be the block diagonal matrix with $\hat{s}_t$ as its block diagonal entries, then we have

$$ \hat{s}_D\hat{s}_D'X_2 = \begin{bmatrix}
(\hat{s}\hat{s}'X_2)_1 \\ \vdots \\ (\hat{s}\hat{s}'X_2)_T
\end{bmatrix} $$

\noindent{}as desired. To accommodate the Hadamard product with $\nu$, we note that

$$\hat{s}'_DX_2 = \begin{bmatrix}
(\hat{s}'X_2)_1 \\ \vdots \\ (\hat{s}'X_2)_T
\end{bmatrix} $$

\noindent{}so we need to take the Hadamard product against a $\nu$ that has been repeated $T$ times and stacked on top of each other. Call this matrix $\nu_D$, then we have that the second term is given at the overall level by

$$\hat{s}_D(\nu_D\odot[\hat{s}_D'X_2]) $$

\noindent{}Unfortunately, the share matrix depends on the value of $\theta_2$ being considered, so it cannot be passed in block diagonal form to a gradient evaluation call. However, there are matrix operations that will convert the share matrix into block diagonal form. Denote $I_R \in \mb{R}^{R \times TR}$ to be the matrix of $T$ identity matrices of size $R$ stacked side by side, and $\iota_R$ to be the block-diagonal matrix with matrices of 1s of size $J_t \times R$ as each block diagonal element. Then we have

$$ \hat{s} I_R = \begin{bmatrix} \hat{s}_1 & \cdots & \hat{s}_1 \\ \hat{s}_2 & \cdots & \hat{s}_2 \\ \vdots & \ddots & \vdots \\ \hat{s}_T & \cdots & \hat{s}_T \end{bmatrix} $$

\noindent{}Once again we have an over-provisioned dense matrix with the correct block diagonal elements, and we use $\iota_R$ to ``slice'' them out.

$$ \hat{s}_D = (\hat{s} I_R) \odot \iota_R $$

\noindent{}$I_R$ can also be used to generate $\nu_D$ through

$$ \nu_D = I_R' \nu $$

\noindent{}As in the case of $\frac{\partial\hat{s}}{\partial\delta}$, we have more computations than an equivalent for loop, although the extra computation is with a sparse matrix, which has well optimized methods. Note that $I_R$ and $\iota_R$ can be pre-computed outside of the GMM objective function. Unsurprisingly, there are many shared intermediate computations between the objective and the gradient evaluations, including the most computationally intensive step of solving for the fixed point $\hat{\delta}$. Algorithm \ref{demand_combined_algo} presents a single function that will share intermediate values used by both function and gradient evaluation, which will reduce the computation time for each iteration of the optimization routine for packages that support this sharing.\footnote{We use \texttt{NLopt} and \texttt{Optax} for quasi-Newton and gradient descent optimization methods respectively, which do support this sharing out of the box. \citet{conlon2020best} use the Python optimization library \texttt{SciPy} that does not explicitly allow for this sharing of intermediate values, but \href{https://github.com/jeffgortmaker/pyblp/blob/master/pyblp/configurations/optimization.py}{their implementation} uses a global cache that saves the objective function and gradient for a given value of $\theta_2$ to achieve equivalent functionality.}

\begin{algorithm}[htbp!]
\DontPrintSemicolon
\KwData{$X_1 \in \mb{R}^{J\times K_1},\ X_2 \in \mb{R}^{J\times K_2},\ Z \in \mb{R}^{J\times z},\ \nu \in \mb{R}^{R \times K_2},\ \mc{S} \in\mb{R}^J,\ \iota_{J/J}, \iota_J$ as in Algorithm \ref{demand_obj_algo}; $\iota_R,\ I_R$ as in Section \ref{demand_grad}.}
\KwIn{$\theta_2, \delta_0, W$ as in Algorithm \ref{demand_obj_algo}.}
\KwOut{$\left[q(\theta),\ \nabla q(\theta_2)\right]$ the objective function and its gradient.}
\Begin{
    $\mb{R}^{J\times R} \ni \mu(\theta_2) \longleftarrow X_2 \Omega_{\theta_2} \nu'$\;
    $\mb{R}^J \ni \hat{\delta} \longleftarrow \lim_{n\to\infty}\delta_n(\delta_0; \mu(\theta_2))$\;
    $\mb{R}^{J \times R} \ni \hat{s}(\hat{\delta};\theta_2) \longleftarrow \exp(\hat{\delta} +^c \mu(\theta_2)) \oslash \left[1 + \iota_J\exp(\hat{\delta} +^c \mu(\theta_2))\right]$\;
    $\mb{R}^{K_1} \ni \hat{\theta_1} \longleftarrow \left[X_1'ZWZ'X_1\right]^{-1}X_1'ZWZ'\hat{\delta}$\;
    $\mb{R}^J \ni \xi \longleftarrow \hat{\delta} - X_1\hat{\theta}_1$\;\;

    \textit{Compute the Jacobian $\frac{\partial \delta}{\partial \theta_2}$}\;
    $\frac{\partial \hat{s}}{\partial\delta}_{i\neq j} \longleftarrow \hat{s}\hat{s}'\odot\iota_J$\;
    $\frac{\partial \hat{s}}{\partial\delta}_{i=j} \longleftarrow rowMeans(\hat{s}\odot(1-\hat{s}))$\;
    $ \hat{s}_D \longleftarrow (\hat{s} I_R) \odot \iota_R $\;
    $ \nu_D \longleftarrow I_R' \nu $\;
    $ \frac{\partial \hat{s}}{\partial\theta_2} \longleftarrow \left(\hat{s}\nu\right)\odot X_2 - \hat{s}_D(\nu_D\odot[\hat{s}_D'X_2]) $\;
    $ \frac{\partial \delta}{\partial \theta_2} \longleftarrow - \left[\frac{\partial \hat{s}}{\partial\delta}\right]^{-1} \frac{\partial \hat{s}}{\partial\theta_2}$\;\;
    
    $q(\theta_2) \longleftarrow \xi'Z W Z'\xi$\;
    $\nabla q(\theta_2) \longleftarrow 2 \frac{\partial\delta}{\partial\theta_2}'Z W Z'\xi$
}
\caption{NFP Objective and Gradient Combined\label{demand_combined_algo}}
\end{algorithm}

\subsection{Uncertainty Quantification}
Using the usual GMM standard errors under similar assumptions to \ref{sec:blpdo_obj}, we have

$$ \Sigma_{\theta} = \left[\frac{\partial \xi(\theta_2)'Z}{\partial \theta} W \frac{\partial Z'\xi(\theta_2)}{\partial \theta}\right]^{-1} \left[\frac{\partial \xi(\theta_2)'Z}{\partial \theta} W \left(Z'\Omega_\xi  Z\right) W \frac{\partial Z'\xi(\theta_2)}{\partial \theta}\right] \left[\frac{\partial \xi(\theta_2)'Z}{\partial \theta} W \frac{\partial Z'\xi(\theta_2)}{\partial \theta}\right]^{-1}$$

where $W$ is the weight matrix used in the second stage of the GMM estimation.This section addresses how to estimate $\frac{\partial Z'\xi(\theta_2)}{\partial \theta}$ without automatic differentiation. By the rules of matrix differentiation, we have

$$ \frac{\partial Z'\xi(\theta_2)}{\partial \theta} = Z'\frac{\partial \xi(\theta)}{\partial \theta} = Z'\begin{bmatrix}\frac{\partial\xi(\theta_2)}{\partial \theta_2} & \frac{\partial\xi(\theta_2)}{\partial \theta_1}\end{bmatrix} $$

This can be written as

$$ \frac{\partial\xi(\theta_2)}{\partial \theta_1} = \frac{\partial}{\partial \theta_1}\delta(\hat{\theta_2}) - X_1\theta_1 = -X_1 $$

\begin{align*}
\frac{\partial\xi(\theta_2)}{\partial\theta_2}
&= \frac{\partial\delta(\theta_2)}{\partial\theta_2} - X_1\frac{\partial \hat{\theta}_1(\hat{\theta}_2)}{\partial \theta_2} = \frac{\partial\delta(\theta_2)}{\partial\theta_2} - X_1\left[X_1'ZWZ'X_1\right]^{-1}X_1'ZWZ'\frac{\partial\delta(\theta_2)}{\partial\theta_2} \\
&= (I-X_1\left[X_1'ZWZ'X_1\right]^{-1}X_1'ZWZ')\frac{\partial\delta(\theta_2)}{\partial\theta_2}
\end{align*}

\noindent{}where the first line in $\frac{\partial\xi(\hat{\theta})}{\partial\theta_2}$ uses the fact that our linear parameters are estimated using one-stage linear GMM. Note that the Jacobian of $\delta$ with respect to $\theta_2$, $\frac{\partial\delta(\theta_2)}{\partial\theta_2}$, is the same as in the gradient of the BLP objective, and is covered in Section \ref{demand_grad}. 

\end{document}